# 2024 California Community Earth Models for Seismic Hazard Assessments Workshop Report

Brad T. Aagaard, U.S. Geological Survey
Scott Marshall, Appalachian State
Sarah Minson, U.S. Geological Survey
Dan Boyd, California Geological Survey
Marine Denolle, University of Washington
Eric Fielding, National Aeronautics and Space Administration
Alice-Agnes Gabriel, University of California at San Diego
Christine Goulet, U.S. Geological Survey
Russell Graymer, U.S. Geological Survey
Jeanne Hardebeck, U.S. Geological Survey
Alexandra Hatem, U.S. Geological Survey
Evan Hirakawa, U.S. Geological Survey
Tran Huynh, Statewide California Earthquake Center
Lorraine Hwang, Computational Infrastructure for Geodynamics
Karen Luttrell, Louisiana State University
Kathryn Materna, University of Colorado
Laurent Montesi, University of Maryland
Michael Oskin, University of California at Davis
Arthur Rodgers, Lawrence Livermore National Laboratory
Arben Pitarka, Lawrence Livermore National Laboratory
Judy Zachariasen, California Geological Survey

# Table of Contents

# Executive summary

The California Community Earth Models for Seismic Hazard Assessments Workshop (https://www.scec.org/workshops/2024/california-community-models, accessed December 16, 2024) was held online on March 4–5, 2024, with more than 200 participants over two days. In this report, we provide a summary of the key points from the presentations and discussions. We highlight three use cases that drive the development of community Earth models, present an inventory of existing community Earth models in California, summarize a few techniques for integrating and merging models, discuss potential connections with the Cascadia Region Earthquake Science Center (CRESCENT), and discuss what "community" means in community Earth models. Appendix B contains the workshop agenda and Appendix C contains a list of participants.

## Key points

- Making community Earth models accessible to users from a variety of technical backgrounds and disciplines is critical. This could be achieved by standardizing metadata, using standard scientific data formats, and providing standard interfaces for accessing each type of Earth model.
- The long-term sustainability of community Earth models requires a community of core developers and contributors with well-defined workflows to incorporate contributions into the models with appropriate attribution.
- The earthquake science community would benefit from consistent definitions of the state of the crust and upper mantle across the various types of community Earth models, extending the spatial coverage, reconciling discrepancies, and developing alternatives consistent with diverse constraints.
- Propagating uncertainty through user applications requires quantitative assessments of epistemic uncertainty and aleatory variability for Earth models, which are lacking for most existing models. These assessments can help prioritize regions for improvement based on uncertainty and seismic hazard and risk.
- Coordination among organizations working on community Earth models across the western United States will improve user accessibility and facilitate the advancement of earthquake science.

# Objectives

The workshop aimed to foster collaboration and establish a strong community around developing and maintaining community Earth models for California. These models (Table 1) describe the state of the Earth's crust and upper mantle, including features like faults, elastic properties, and stress, which are crucial for understanding earthquake processes. The workshop emphasized the importance of "community" in these models, meaning they are publicly accessible and regularly updated through ongoing collaboration with the scientific community. Community Earth models are critical inputs for various earthquake hazard assessment studies. Several organizations, such as the Statewide California Earthquake Center (SCEC), the U.S. Geological Survey (USGS), and the Cascadia Regional Earthquake Science Center (CRESCENT), are creating models that cover various regions of California. This workshop marks a crucial step toward building a collaborative and sustainable framework for developing and utilizing community Earth models for the entire state of California.

The workshop objectives were to

1. **Form a collaborative community** by bringing together researchers and stakeholders interested in developing and updating Earth models for California.
2. **Promote sustainability** to ensure the long-term development and maintenance of these models.
3. **Embrace multiscale approaches** by considering Earth models at different scales for a comprehensive understanding.
4. **Assess existing models** by evaluating the spatial coverage to identify gaps and regions where models could be merged.
5. **Define short and long-term goals** to outline priorities and milestones for future development.

**Table 1**: Common types of community Earth models.

| Name | Description |
| --- | --- |
| Geologic model | Three-dimensional model of geologic units |
| Fault geometry model | Three-dimensional model of the geometry of fault surfaces |
| Seismic wave speed (velocity) model | Model of elastic material properties and often intrinsic attenuation, usually as a function of depth or in three dimensions |
| Rheology model | Model of viscoelastic or elastoplastic material properties, usually a table mapping bulk constitutive models to geologic units |
| Thermal model | Model of temperature, usually as a function of depth or in three dimensions |
| Stress model | Model of differential or absolute stress |
| Geodetic model | Displacements at discrete points on Earth's surface, usually velocities defining long-term motion associated with plate tectonics |

## Roles in developing and maintaining community Earth models

Scientists participate in the process of developing and maintaining community Earth models in a variety of ways. An individual scientist may simultaneously serve multiple roles, such as contributor and user.

**Developer or maintainer**: A person who assembles model components from data sources or integrates such contributions from others.

**Contributor**: A person who contributes features to models for integration by a developer or maintainer.

**User**: A person who uses a model in a scientific study, outreach information, or other application.

**Spectator**: A person interested in exploring or learning how models are developed and used.

# Use cases for California community Earth models

## Use case 1: Advancing earthquake science

Numerical models are essential for advancing our understanding of the physical processes underlying the dynamics of Earth's crust and upper mantle. Community Earth models provide critical constraints on the geologic structure, rheologic behavior, and deformation rates in such models. We can address critical scientific questions by integrating community Earth models in physics-based simulations. Here, we highlight earthquake simulations using dynamic (spontaneous) rupture models and how they leverage Earth models to address scientific questions related to earthquake rupture dynamics, seismic wave propagation, earthquake recurrence patterns, rupture extent, and interseismic deformation.

Modeling the dynamics of complex multi-segment fault systems such as the 2019 magnitude 6.4 Searles Valley and magnitude 7.1 Ridgecrest, California, earthquakes (Taufiqurrahman and others, 2023) and the 2023 Kahramanmaraş, Turkey, doublet (Gabriel and others, 2023; Jia and others, 2023), demonstrates how regional structures, ambient stresses, and co-seismic fault interactions affect the dynamics and timing of seismic sequences. These simulations also reveal the spectral fingerprints of localized impulsive ground motion in near-field waveforms arising from fault roughness and topography (Schliwa and Gabriel, 2023). Incorporating lower rigidities in a fault zone influences ground motion far beyond the immediate vicinity of the rupturing faults. Furthermore, spatially heterogeneous model parameters from community Earth models produce realistic hypocenter variability in models of earthquake sequences.

The digital twin concept, in which rich datasets and high-fidelity Earth models drive numerical models for detailed hypothesis testing, encapsulates the potential benefits behind community Earth models. Likewise, community Earth models can provide the datasets needed for training machine-learning reduced-order models that support rapid hazard assessment and warning systems, such as near real-time ground-motion estimates (Rekoske and others, 2023).

## Use case 2: Improving seismic hazard assessments

Standard methods for seismic hazard assessment (for example, Gerstenberger and others, 2023; Meletti and others, 2021; Petersen and others, 2024) involve several community Earth models (fault geometry models, geodetic models, and seismic wave speed models), and more novel techniques can require additional community Earth models (rheology models, thermal models, or stress models). In this section, we focus on the hazard associated with ground shaking, but similar remarks also apply to the hazard related to displacement across faults.

Traditional seismic hazard assessment for ground shaking, whether it is deterministic or probabilistic, requires two main components: an earthquake rupture forecast and an estimation of ground shaking for those earthquake ruptures (for example, Gerstenberger and others, 2023; Petersen and others, 2024). Earthquake rupture forecasts often rely on community Earth models to describe the fault geometry and geodetic deformation to compute fault slip rates (for example, Field and others, 2014; Field and others, 2023). The ground-motion models used to estimate shaking (such as Bozorgnia and others, 2014; Bozorgnia and others, 2022) can leverage community Earth models to define site conditions, such as the near-surface material properties (often parameterized by the time-averaged shear wave speed in the top 30 m) and the depths of sedimentary basins (for example, Bradley and others, 2022; Moschetti and others, 2024). Community Earth models are increasingly used in seismic hazard assessments to determine the depths of sedimentary basins. However, their application tends to be limited by the resolution of seismic wave speed models and the challenges in predicting ground-motion amplitude based on these depths (Ahdi and others, 2024).

Simulation-based seismic hazard assessment quantifies ground shaking using numerical models of earthquake rupture and seismic wave propagation (for example, Graves and others, 2010; Jordan and others, 2018). Alternative approaches for these numerical models have varying levels of complexity. The more straightforward approaches rely on the same community Earth models as traditional seismic hazard assessment. Approaches using more sophisticated methods, such as dynamic or quasi-static spontaneous rupture simulations, can leverage community Earth models describing the stress field, fault, bulk rheologies, and seismic wave speed models.

### Validation and consistency

Validation of community Earth models used in seismic hazard assessment is critical. In some cases, individual models can be validated against observations, but in many cases, validation requires integrating multiple models to make a comparison with observations possible. For example, seismic wave speed models defining bulk elastic properties can be validated using ground-motion simulations. Wave propagation simulations directly test a seismic wave speed model for small earthquakes using ground-motion records (some examples in California include Hirakawa and Aagaard, 2022; Kim and others, 2010; Lee and others, 2014; Olsen and Mayhew, 2010; Taborda and Bielak, 2014). Validation for large earthquakes requires additional information, such as fault geometry and a model of the rupture propagation, which often come from other types of community Earth models (for example, Graves and Aagaard, 2011). Features should be consistent across different types of community Earth models. For example, when a fault surface separates two geologic units with different rigidities, we want that surface in a fault geometry model to align with the change in rigidity in the seismic wave speed model.

### Improving community Earth models

Improving community Earth models can increase the accuracy and precision of seismic hazard assessments. Improving the horizontal and vertical resolution of sedimentary basins in seismic wave speed models could facilitate improving the accuracy and precision of basin response in ground-motion models. For example, seismic wave speed models derived from seismic tomography tend to have poor resolution at depths less than 1 km, which is critical for resolving the geometry of sedimentary basins. Additionally, long-term accumulation of slip on faults tends to generate a damage zone around the fault with a reduced rigidity, which affects the radiated seismic waves and the local distribution of shaking (examples in California include Catchings and others, 2016; Li and others, 1994; Spudich and Olsen, 2001). Most seismic wave speed models lack finite-width fault zones with reduced rigidity (Boyd and Shah, 2018; Aagaard and Hirakawa, 2021b; SCEC, 2021 (CVM-H); SCEC 2022a (CCA06); SCEC, 2022b (CVM-S4); Doody and others, 2023).

Seismic hazard assessments involve detailed characterization of the uncertainties arising from all model components. As a result, they need to consider uncertainty quantification for individual community Earth models and the development of suites of viable alternative models to capture both the epistemic uncertainty and aleatory variability. As discussed in the section on the inventory of existing models in California, in many cases, we have a small number of models covering any given location and need quantitative estimates of uncertainty.

## Use case 3: Carbon sequestration and community Earth models

The community Earth models discussed in this report target applications related to seismic hazard assessment. Such models provide important information for use in other scientific studies that involve the state of Earth's crust, and the broader community could benefit from coordinating and collaborating in the development of Earth models. In this use case, we highlight carbon sequestration as an example of how community Earth models can be useful outside of seismic hazard applications. Identifying sites suitable for $CO_2$ injection requires detailed knowledge of the geologic structure and the stress field. California Geological Survey's geologic modeling efforts will transition into the newly established Geologic Carbon Sequestration Group mandated by California Senate Bill 905 (SB-905, 2022). A key objective of the new group will be to ensure stable, long-term storage of $CO_2$ and monitor for potential hazards. In addition to community Earth models of fault and geologic structure, this work may also leverage community Earth models for elastic properties and stress.

These additional applications of community Earth models beyond seismic hazard assessment present opportunities for collaboration. For example, seismic hazard assessments and carbon sequestration studies benefit from improving semi-automated methods for deriving fault geometry from seismicity data (for example, Plesch and others, 2020; Riesner and others, 2017) and a better understanding of the stress field.

# Overview and inventory of existing community Earth models

As part of the workshop, we compiled a list of various existing Earth models. With an emphasis on future model development, we focus on well-known or widely used models and models under development rather than assembling a comprehensive list. Appendix A includes figures showing the geographic extent of each model.

## Geologic models

Three-dimensional (3D) geologic models describe the distribution of geologic units and structures, such as fault geometry and sedimentary basis. Geologic models underlie some other types of community models, such as seismic wave speed models and rheology models. Geologic models (Figure 1 and Table 2) vary in focus, scale, and resolution. The USGS National Crustal Model (Boyd and Shah, 2018) provides a coarse representation of the 3D geologic structure covering California. Still, only about one-third of the state is represented in one or more detailed models covering southern California, the central Coast Ranges, the San Francisco Bay region, or the Central Valley. The northern Coast Ranges, Klamath Mountains, Sierra Nevada, Walker Lane, and northeastern California are almost entirely without detailed 3D models, although the Eel River Basin is a notable exception.

3D geologic models fall into two principal types: crustal-scale models (for example, Boyd and Shah, 2018; Jachens, 2006; SCEC, 2021) and basin models (Magistrale and others, 1996). Crustal-scale models are often used in seismic hazard studies (for example, Graves and others, 2010; Jordan and others, 2018) or oil and gas assessments (for example, Scheirer, 2007). These models reach down to the base of seismicity (10–15 km) or the base of the crust, and many incorporate some description of the top of the mantle. They tend to be of regional scale and have a generalized stratigraphic framework and simplified (though still complex) fault model (for example, Boyd and Shah, 2018; Jachens, 2006). Deep crustal composition and structure are inferred from sparse, exhumed crustal sections, tectonic history, and geophysical data such as seismic wave speeds, gravity, and magnetic properties.

In contrast, basin models are often developed for groundwater studies (for example, Traum and others, 2022). These models are generally much shallower, focusing on the upper 3–5 km of the crust. They tend to be more detailed because they incorporate a smaller volume and span depths with more direct observation (for example, petroleum or water wells). Basin models also consider various types of units, including stratigraphic, lithologic, and hydrogeologic. Only a few areas in the state have multiple models, principally the San Joaquin Valley (for example, Gooch, 2022; Scheirer, 2007) and basins within the central Coast Ranges and San Francisco Bay region, where smaller basin models (for example, Cromwell and others, 2024; Sweetkind and Faunt, 2024) reside within regional upper crustal models (Jachens, 2006).

In addition to differences in content, detail, and scope, the models vary in formulation. Some models are constructed from vector surfaces (Jachens, 2006), whereas others are built from raster grids (Boyd and Shah, 2018). Model components are disseminated in a variety of file formats. Some formats, such as EarthVision fault trees, are software-specific, whereas others are more generic, such as text files with surface triangulations or comma-separated values.

Geologic models are interrelated with fault geometry models, which we discuss in the next section. Geologic models often incorporate significant faults that are no longer active, whereas fault geometry models tend to focus on Quaternary-active faults. Both types of models tend to represent faults as surfaces. However, geologic models may include multiple surfaces for a single fault in areas with broad fault zones that incorporate substantially different geologic units, such as uplifted lenses of basement rock within flower structures. Ideally, the fault surfaces in both geologic models and fault geometry models would be the same, so coordination between the developing groups is essential.

Some potential goals and milestones for community development of geologic models in California include the following:

1. Create a community of contributors and developers based on open-source tools with well-defined workflows for incorporating contributions into models.
2. Assess discrepancies among models where they overlap within California, reconciling differences and documenting alternative representations where appropriate.
3. Where feasible, incorporate details from smaller high-resolution models into larger regional models. For other situations, develop methods and tools to embed smaller, high-resolution models within larger models.
4. Leverage evaluation of community Earth models derived from geologic models to improve the underlying geologic models.

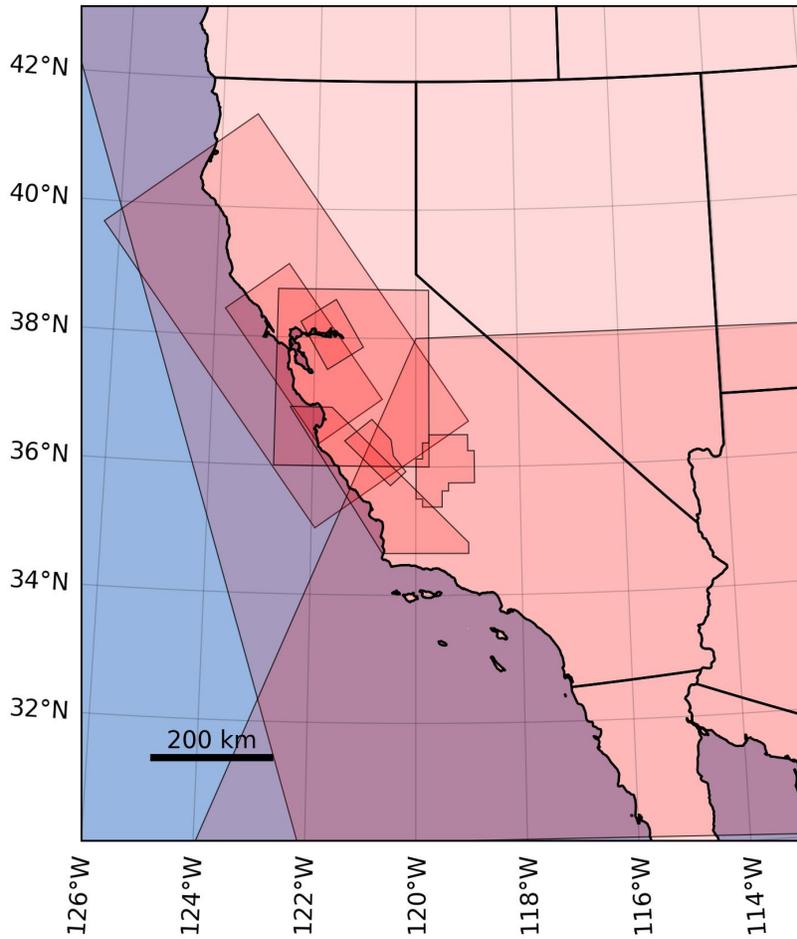

**Figure 1**: The shaded, semi-transparent polygons show the geographic coverage of the geologic models in Table 2.

Table 2: Inventory of existing geologic models.

| Name | Authors | Description | Reference | Status | Bounding Box |
|---|---|---|---|---|---|
| USGS NCM | Boyd and others | Geophysical model of the conterminous United States | (Boyd and Shah, 2018) | Updates | (22.02, 48.40, -129.70, 63.67) |
| USGS SF: Regional domain | Jachens and others | 3D geologic model of the San Francisco Bay region | (Jachens, 2006) | Updates | (35.04, 41.46, -126.32, -118.98) |
| USGS SF: Detailed domain | Jachens and others | 3D geologic model of the San Francisco Bay urban region | (Jachens, 2006) | Updates | (36.35, 39.14, -123.80, -120.66) |
| SCEC GFM | Oskin and others | Geologic block model based on 1D lithologic columns | (Oskin, 2024) | In progress | (30.00, 38.00, -124.00, -112.00) |
| Medwedeff SF geologic model | Medwedeff | Compilation of map-based data, including geology, geophysical, petroleum well, and seismic datasets | | In progress | (36.00, 38.75, -122.75, -119.75) |
| Southern San Joaquin GFM | Gooch and others | Geologic framework model from the surface to crystalline basement | (Gooch, 2022) | In progress | (35.38, 36.50, -120.00, -118.88) |
| Geologic model of the Sacramento-San Joaquin River delta | Graymer and others | Upper crustal 3D geologic model, including faults and principle stratigraphic packages | | Developed | (37.51, 38.60, -122.29, -121.04) |
| Geologic Model of the Central Coast Ranges | Graymer and McFaul | Upper crustal 3D geologic model, including faults and principle stratigraphic packages | | In progress | (34.65, 36.92, -122.46, -119.00) |
| Geologic Model of the San Andreas Fault Zone | Roberts and others | 3D geological model derived from geologic mapping, potential field geophysics, and petroleum well logs | | Developed | (35.71, 36.74, -121.39, -120.20) |
| USGS 3D Geologic Model Inventory | Many models with multiple authors within California | Previously published (2004–2022) USGS 3D geological models | (Sweetkind and Zellman, 2022) | Updates | multiple models statewide |

The bounding box is (minimum latitude, maximum latitude, minimum longitude, maximum longitude) in degrees in the WGS84 horizontal datum.
1D: One dimensional
3D: Three dimensional
GFM: Geologic Framework Model
NCM: National Crustal Model
SCEC: Statewide California Earthquake Center
SF: San Francisco
USGS: U.S. Geological Survey
Updates: Model updates are anticipated on a regular or irregular schedule
Developed: Model has been developed but is not yet published
In progress: Model is under initial development

# Fault geometry models

Fault geometry models define the geometry of fault surfaces, including the strike, dip, and depth extent. Some include detailed 3D representations, whereas others include limited detail, often presented as rectilinear surfaces. Two main fault geometry models (Figure 2 and Table 3) span California: the SCEC community fault model (CFM) and the National Seismic Hazard Model fault sections database (NSHM FSD, Hatem and others, 2023). The SCEC CFM (Plesch and others, 2007; Plesch and others, 2023) has nonplanar 3D surfaces.

The SCEC CFM provides discretized surfaces for more than 400 faults, incorporating available data to constrain subsurface geometries at various resolutions. The fault surfaces can be quite detailed and are used to routinely determine causative faults of earthquakes within the model region (Evans and others, 2020). The SCEC CFM benefits from a plethora of data, particularly industry subsurface data, that illuminate the 3D fault geometry.

The NSHM FSD provides a comprehensive summary of faults across 12 western U.S. states likely capable of hosting M6.5+ earthquakes. The NSHM FSD was constructed for probabilistic seismic hazard analysis. The fault geometry is "2.5D," meaning each fault has a relatively simple fault trace (minimum discretization size is 1 km) with a dip angle, dip direction, and lower and upper seismogenic depths as attributes. Surfaces can be constructed by extruding fault traces using these attributes, but the FSD was not built as a 3D model. For example, some faults cross each other in the subsurface in non-physical ways. This model builds upon the Uniform California Earthquake Rupture Forecast version 3 fault data (Dawson, 2013).

Fault geometry models require criteria for including or excluding a given fault. For example, not all faults in the SCEC CFM are included in the NSHM FSD. The use case for the NSHM FSD dictates simpler faults than what is depicted in the SCEC CFM. However, the SCEC CFM provides a set of faults for potential inclusion in the NSHM FSD and an essential basis for developing simplified representations.

Additional fault databases exist in California, including the Quaternary Fault and Fold Databases (U.S. Geological Survey, 2022) hosted by the California Geological Survey and the USGS. These are considered fault inventories and typically contain little information about fault geometry at depth. The surface traces of the faults in these inventories are highly detailed and represent field and remote observations of fault scarps and many geomorphic indicators. Fault inventories catalog what is known about each fault in a given area and provide a good starting point for users. Additionally, the Cascadia Region Earthquake Science Center (CRESCENT) aims to improve the NSHM FSD fault geometries in northern California by incorporating additional detail and expanding farther offshore.

Improving existing fault geometry models is essential for developing community Earth models across California. CRESCENT's CFM, which will extend down to the Mendocino Triple Junction, complements ongoing efforts by SCEC, which are focused south of the Mendocino Triple Junction. The USGS plans to overhaul the subsurface representation of the NSHM FSD in the coming years, using techniques and tools used in the SCEC CFM and fault geometry models in other countries, such as New Zealand (Seebeck and others, 2024). Common infrastructure, such as the SCEC CFM web browser viewer, and standardizing formats could facilitate coordination among these groups.

Some potential goals and milestones for community development of fault geometry models in California include the following:

1. Prioritize filling in gaps in the geographic coverage of models.
2. Disseminate models in standard scientific formats with standard metadata.
3. Develop methods for quantifying the uncertainty in fault geometry, especially extrapolating fault traces with depth.
4. Standardize the process for developing consistent complex and simplified representations of fault geometry.
5. Develop tools for generating stochastic realizations of representative 3D fault geometry in regions where models are incomplete.

**Table 3**: Inventory of existing fault geometry models

| Name | Authors | Description | Reference | Status | Bounding Box | Methodology |
|---|---|---|---|---|---|---|
| NSHM23 | Hatem and others | Fault geometries in 2023 NSHM | (Hatem and others, 2023) | Published | (28.54, 50.18, -126.38, -102.92) | 2.5D representation of faults |
| SCEC community fault model | Plesch and others | 3D representation of active faults in southern California | (Plesch and others, 2007; Plesch and others, 2023) | Updates | (31.70, 37.30, -121.80, -114.90) | Fully 3D representations |
| Quaternary faults offshore of California | Walton and others | Database of faults offshore California from the U.S.-Mexico border to Cape Mendocino | (Walton and others, 2020) | Updates | (32.30, 40.30, -124.50, -117.10) | Quaternary faults offshore of California |
| USGS Quaternary Fault Database | Numerous USGS contributors | Interactive 2D map of Quaternary active faults | (U.S. Geological Survey, 2022) | Published | Entire U.S. | 2D Field and Geologic map investigations and other methods |

The bounding box is (minimum latitude, maximum latitude, minimum longitude, maximum longitude) in degrees in the WGS84 horizontal datum.

2D: two dimensional

2.5D: three-dimensional model with fault geometry based on fault surface traces and fault dip angle

3D: three dimensional

NSHM: National Seismic Hazard Model

SCEC: Statewide California Earthquake Center

USGS: U.S. Geological Survey

Updates: Model updates are anticipated on a regular or irregular schedule.

Published: Model is published, but no updates are anticipated.

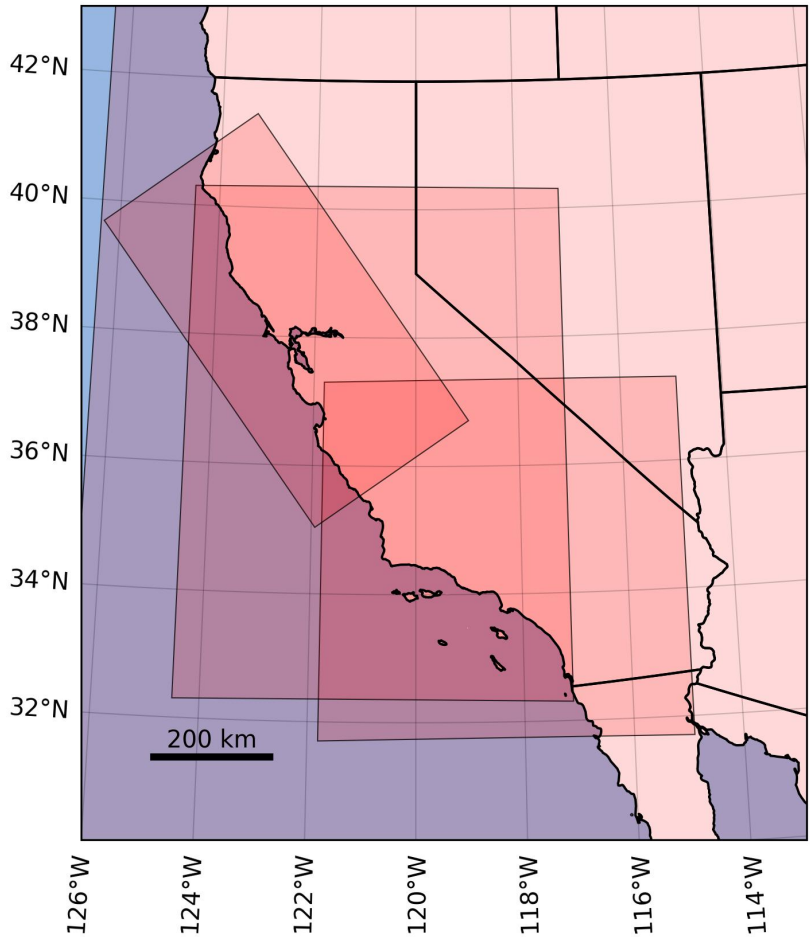

**Figure 2**: The shaded, semi-transparent polygons show the geographic coverage of the fault geometry models in Table 3.

## Seismic wave speed models

Many 3D seismic wave speed models, often called seismic velocity models, span large regions of California. These models usually include density, intrinsic attenuation, and P- and S-wave speeds. The USGS National Crustal Model (NCM, Boyd and Shah, 2018) covers the western United States. There is substantial overlap among models along the San Andreas Fault (Figure 3). Some high-resolution models cover small regions, such as portions of the Los Angeles basin. The northeast portion of California has the least coverage.

The models were constructed using a variety of techniques (Table 4). The four most common methods used to build the models include (1) travel-time tomography, (2) waveform tomography, (3) ambient noise tomogrpahy, and (4) assigning elastic properties to geologic units in 3D geologic models (rule-based models). The discretization and scale of the models depend on the construction technique and data. The discretization size of the tomography-based models generally reflects the resolution of the inversion. The highest-resolution regional-scale tomography models are discretized at scales of about 500 m. Some of the coarse-resolution models have horizontal discretizations of 5–30 km. The rule-based models have the highest resolution discretization, commensurate with the continuous functions used to assign the elastic properties as a function of space.

The resolution of the models varies considerably in the top 1 km. Some models have artificially high P- and S-wave speeds outside their "core" region, where elastic properties are better constrained. Many models that rely on tomography alone do not resolve variations in elastic properties in the top kilometer. The SCEC Unified Community

Velocity Model (UCVM) software (Small and others, 2017) can apply generic shallow elastic properties that are tied to $V_{S30}$ (time average shear wave speed in the top 30 m).

The "community" aspect of the seismic wave speed models could be improved. Only a few models have been updated since they were first published, and only a few have been archived with version numbers.

Some potential goals and milestones for community development of seismic wave speed models in California include the following:

1. Evaluate models using the same dataset(s), which could include recorded ground motions from moderate earthquakes and gravity data.
2. Identify and resolve discrepancies among models while identifying sources of such discrepancies (for example, differences arising from the data used to construct the models or from the construction techniques themselves).
3. Improve the resolution for depths less than 1 km over broad regions, especially in urban areas.
4. Develop techniques for quantifying uncertainties in the elastic properties and earthquake ground-motion metrics consistent with propagating uncertainties in user applications.
5. Improve the consistency of models with other types of community Earth models, such as including contrasts in elastic properties across geologic units where appropriate and sedimentary basin geometry consistent with geologic models.

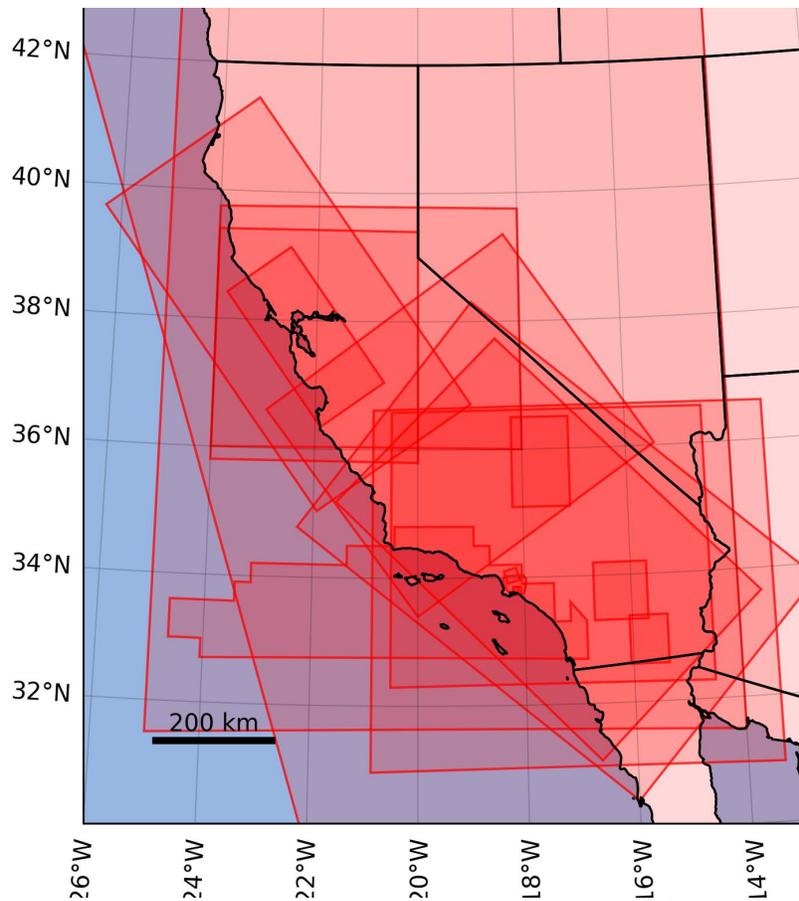

**Figure 3**: The shaded, semi-transparent polygons show the geographic coverage of the seismic wave speed models in Table 4.

Table 4: Inventory of existing seismic wave speed models

| Name | Authors | Description | Reference | Status | Bounding Box | Methodology | Min grid spacing | Availability |
|---|---|---|---|---|---|---|---|---|
| SCEC CVM-S4 | Kohler and others | Model of southern CA focused on the LA basin | (SCEC 2022b) | Published | (31.10, 37.73, -121.57, -113.57) | Geology+rules | 0 m | UCVM |
| SCEC CVM-S4.26.M01 | Lee and Chen | Update to SCEC CVM-S4 using waveform tomography | (SCEC, 2022c) | Published | (30.45, 38.30, -122.30, -112.52) | Geology+tomography | 500 m H, 0 m V (Ely GTL) | UCVM |
| SCEC CVM-H | Shaw and others | Model of southern CA focused on basins | (SCEC, 2021) | Updates | (30.96, 36.61, -120.86, -113.33) | Geology+tomography | 200 m H, 100 m V | UCVM |
| SCEC CCA06 | Lee and others | Model of central CA | (SCEC, 2022a) | Published | (33.40, 39.35, -122.95, -115.45) | Waveform tomography | 500 m | UCVM |
| USGS SF-CVM Detailed domain | Aagaard and Hirakawa | Model covering the SF Bay urban region | (Aagaard and Hirakawa, 2021b) | Updates | (36.35, 39.14, -123.80, -120.66) | Geology+rules | 100 m H, 25 m V | ScienceBase |
| USGS SF-CVM regional domain | Aagaard and Hirakawa | Model covering the greater SF Bay region | (Aagaard and Hirakawa, 2021a) | Updates | (35.04, 41.46, -126.32, -118.98) | Geology+rules | 200 m H, 50 m V | ScienceBase |
| USGS NCM | Boyd and Shah | Model covering the western United States | (Boyd and Shah, 2018) | Updates | (22.02, 48.40, -129.70, -63.67) | Geology+rules | 1 km H, 0 m V | USGS web service |
| CANVAS | Doody and others | California-Nevada Adjoint Simulations (CANVAS) Model | (Doody and others, 2023; Doody, 2023) | Published | (31.50, 43.00, -125.00, -114.00) | Waveform tomography | 5 km H, 1 km V | |

| Name | Authors | Description | Citation | Status | Bounding box | Method | Resolution | Repository |
|---|---|---|---|---|---|---|---|---|
| ALBACORE | Bowden and others | Model of region offshore southern CA | (Bowden and others, 2016) | Published | (32.70, 34.80, -124.65, -116.85) | Ambient noise | 30k m H, 1 km V | UCVM |
| SSIP Imperial | Ajala and others | Model for Imperial Valley | (Persaud, 2022) | Published | (32.60, 33.36, -116.05, -115.34) | Travel time tomography | 1 km | UCVM |
| SSIP Coachella | Ajala and others | Model for Coachella Valley | (Persaud, 2022) | Published | (33.30, 34.20, -116.70, -115.70) | Travel time tomography | 1 km H, 500 m V | UCVM |
| Li and Ben-Zion | Li and Ben-Zion | Multi-scale seismic imaging of Ridgecrest, CA | (Li and Ben-Zion, 2024) | Published | (35.10, 36.50, -118.20, -117.10) | Waveform inversion and ambient noise | 150 m | Zenodo |
| Zhang and Ben-Zion | Zhang and Ben-Zion | Merged multiscale seismic wave speed models for southern CA | (Zhang and Ben-Zion, 2024b) | Published | (32.29, 36.58, -120.51, -114.51) | Sparse dictionary learning | 3 km H, 1 km V | Zenodo |
| Castellanos LB P-wave | | P-wave model of Long Beach, CA | (Castellanos, 2019) | Published | (33.75, 33.85, -118.21, -118.12) | Ambient noise | 70 m | Caltech DATA |
| Castellanos LB S-wave | | S-wave model of Long Beach, CA | (Castellanos, 2021) | Published | (33.73, 33.85, -118.21, -118.06) | Ambient noise | 100 m H, 12.5 m V | Caltech DATA |
| Jia LAS1 | | Model of the LA basin | (Jia, 2020) | Published | (33.73, 34.03, -118.40, -117.97) | Ambient noise | 90 m H, 75 m V | Caltech DATA |
| Muir NE LA Basin | | Model of the northeastern LA basin | (Muir and others, 2021) | Published | (33.91, 34.15, -118.38, -118.09) | Love wave dispersion and amplification | 220 m H, 120 m V | Caltech DATA |
| Guo and Thurber | Guo and others | Model for SF Bay | | In progress | (35.80, 39.40, -124.00, -120.00) | Travel time tomography | | |
| CENOCA_AWT | Rodgers and others | Model for central and northern CA | | In progress | (36.00, 39.75, -124.00, -118.00) | Waveform tomography | | |

The bounding box is (minimum latitude, maximum latitude, minimum longitude, maximum longitude) in degrees in the WGS84 horizontal datum.

The minimum grid spacing includes horizontal (H) and vertical (V) values when they are not the same.
ALBACORE: Asthenospheric and Lithospheric Broadband Architecture from the California Offshore Region Experiment
CA: California
CCA: Central California
CVM: Community Velocity Model
EMC: EarthScope Earth Model Collaboration (NetCDF; not for large files)
GTL: Geotechnical layer
LA: Los Angeles, California
LB: Long Beach, California
NCM: National Crustal Model
SCEC: Statewide California Earthquake Center
SF: San Francisco, California
SSIP: Salton Sea Imaging Project
UCVM: SCEC Unified Community Velocity Model software
USGS: U.S. Geological Survey
Updates: Model updates are anticipated on a regular or irregular schedule.
Published: Model is published, but no updates are anticipated.
In progress: Model is under initial development.

# Rheology models

Rheology is the relationship between the stress experienced by a body and the resulting deformation. Because rheology describes a response, it cannot be directly probed but must be predicted based on a theoretical treatment or inferred from models of stress and either strain or strain rates (for example, Burov, 2011; Morrison, 2001; Rutter and Brodie, 1991). Such models are typically parameterized using elastic moduli or viscosity. Elastic properties can be parameterized by density and P- and S-wave speeds in seismic wave speed models. However, high-frequency loading by seismic waves might elicit a different response than a steady, long-term load (Cheng and Johnston, 1981; Shen and others, 2024; Simmons and Brace, 1965).

Permanent deformation can accumulate from brittle failure or ductile flow. Brittle failure may be parameterized using a maximum rock strength or yield stress (for example, Jaeger and others, 2007). However, brittle failure is typically associated with faults and surrounding damage zones (Choi and others, 2016; Kim and others, 2004). Distributed brittle failure is possible, but many questions remain about the physical mechanisms (Ben-Zion and Dresen, 2022; Pan and Wen, 2015). Thus, regional-scale, spatially continuous models of brittle deformation parameters are rare (Jacquey and Cacace, 2020; Karrech and others, 2011; Manaker and others, 2006). Ductile flow occurs at a regional scale but only at depths where the temperature and pressure are high enough to suppress brittle failure. The depth at which ductile creep takes place limits the possibilities of direct observations to an effective viscosity linked to postseismic creep (Savage, 1990). However, further detection methods, such as climate forcing, are being developed.

Two rheology models cover most of southern California (Figure 4 and Table 5). The SCEC Community Rheology Model (CRM, Hearn and others, 2020) estimates ductile flow law parameters. Its current version considers a single deformation mechanism, dislocation creep. Each set of parameters is linked to a geologic unit, not a geographical location. Thus, the CRM does not have an intrinsic geographical extent or a resolution. It was designed to function with SCEC's Geologic Framework (Table 2) but may be applied to similarly defined geologic models. The CRM is based on selected flow laws for eight minerals describing the geologic framework's various geologic units. It can be expressed as a field of effective viscosity when coupled with a geologic model, a temperature model, and a strain rate or stress model. Therefore, the CRM is distributed with SCEC's Geologic Framework and Community Thermal Model (CTM, Thatcher and others, 2020).

An alternative approach, included in the model by Shinevar and others (2018), converts seismic wave speed into viscosity with the intermediary of a petrological model and a flow law mixing model similar to that used in the CRM. Shinevar and others (2018) fit systematic P- and S-wave speeds and viscosity calculations for various mineral assemblages at temperatures, pressures, water fugacity, and strain rates relevant to the lower crust. Then, they generated a temperature model and used elastic properties in a community seismic wave speed model to convert surface strain rate measurements into a viscosity model.

Some potential goals and milestones for community development of rheology models in California include the following:

1. Extend the application of the SCEC Community Rheology Model to community geologic models other than the SCEC Geologic Framework, incorporating additional constitutive models as necessary to represent the expanded diversity of geologic units and behavior.
2. Evaluate the SCEC Community Rheology Model in multiple applications and prioritize regions and features for improvement.
3. Assess alternative approaches for developing community rheology models and integrating rheology with other types of community Earth models.

Table 5: Inventory of existing rheology models.

| Name | Authors | Description | Reference | Status | Bounding Box |
|---|---|---|---|---|---|
| SCEC CRM | Hearn and others | Flow law parameters based on mineral assemblage and dislocation creep | (Hearn and others, 2020; Hearn and others, 2022) | Updates | Not applicable |
| California Thermal Model and Rheology Model | Shinevar and others | Viscosity deduced from velocity, temperature, and strain rate | (Thatcher and others, 2020; Thatcher and others, n.d.) | Updates | (32.00, 37.00, -121.00, -114.00) |

The bounding box is (minimum latitude, maximum latitude, minimum longitude, maximum longitude) in degrees in the WGS84 horizontal datum.

CRM: Community Rheology Model

SCEC: Statewide California Earthquake Center

Updates: Model updates are anticipated on a regular or irregular schedule.

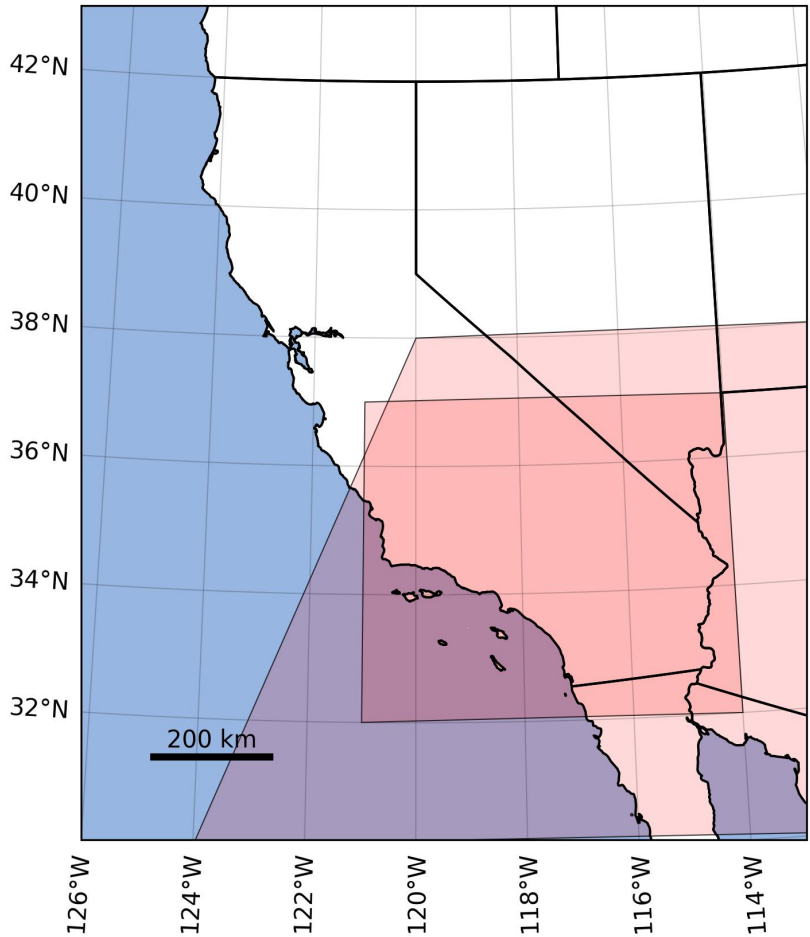

**Figure 4**: The shaded, semi-transparent polygons show the geographic coverage of the rheology models in Table 5.

## Thermal models

Temperature is important for understanding rheology and geothermal resources. It can be directly measured in boreholes near the surface and extrapolated to deeper depths using a theoretical foundation. The simplest models assume steady-state conduction. They depend on heat conductivity and production, which must be inferred from geologic information. These assumptions lead to uncertainty in temperature models, further compounded by the variability of the heat flow measurements themselves, which arises from various near-surface phenomena. Several thermal models span portions of California or the western United States (Figure 5 and Table 6).

To circumvent these difficulties, Thatcher and others (2020) define various "heat flow provinces" and propose a "geotherm" for each province that relates temperature and depth. The geotherms extend into the mantle, where the depth of the lithosphere-asthenosphere boundary further constrains them. In a few locations, steady-state temperature profiles are untenable, and this model considers the temporal adjustment of temperature to a sudden change in lithosphere thickness induced by delamination. In the oceanic domain, the model uses a transient cooling model parameterized by plate age. This model is distributed as SCEC's Community Thermal Model (CTM).

Shinevar and others (2018) follow an alternative approach using smoothed and interpolated heat flow to capture lateral temperature variations within heat flow provinces. However, this model ignores regional geological variations. Shinevar and others (2018) further simulate thermal diffusion over 5 million years to remove the sharp temperature contrasts that appear at the boundary of the heat flow provinces. This model and the CTM illustrate choices between model generality and geologic specificity and are distributed together. Neither has an underlying grid.

Other temperature models have been developed for a broader region that encompasses California. The National Crustal Model contains a 3D temperature model defined on a 0.5° grid (Boyd and Shah, 2018). In the continental crust, the model consists of steady-state one-dimensional temperature profiles constrained by Earth's surface temperature gradient and the depth and temperature at the Mohorovičić discontinuity (Moho), deduced from Pn measurements (variations in travel time for P waves bottoming in the uppermost mantle or emitted from sources in the uppermost mantle). Furthermore, the model assumes that heat production and thermal conductivity decrease exponentially with depth. These temperature profiles continue into the mantle, where a mantle adiabat and melting limit temperature. In the oceans, the temperature is given by a half-space cooling model based on seafloor age and the temperature at the top of the crust and the convecting mantle. The model further includes the effect of surface temperature over the last 5 million years but not geological complexity or recent tectonic events. Blackwell and others (2011) presented a broadly similar model, with additional complexity in the upper crust, motivated by energy resource estimates throughout the United States.

Some potential goals and milestones for community development of thermal models in California include the following:

1. Identify discrepancies among models where they overlap, reconcile different models, and document alternative representations as appropriate.
2. Evaluate the models in multiple applications and prioritize regions for improvement.
3. Assess alternative approaches for developing community thermal models and integrating thermal models with other types of community Earth models.

Table 6: Inventory of existing thermal models.

| Name | Authors | Description | Reference | Status | Bounding Box |
|---|---|---|---|---|---|
| SCEC CTM | Thatcher and others | Temperature profiles associated with heat flow provinces | (Thatcher and others, 2020; Thatcher and others) | Updates | (29.40, 37.70, -123.60, -112.20) |
| California Thermal Model and Rheology Model | Shinevar and others | Temperature profiles based on surface heat flow, steady-state, and temperature profiles. | (Hearn and others, 2020; Shinevar and others, 2018) | Updates | (32.00, 37.00, -121.00, -114.00) |
| USGS Thermal Model for Seismic Hazard Studies | Boyd | Grids in support of the U.S. Geological Survey Thermal Model for Seismic Hazard Studies | (Boyd, 2019) | Updates | (22.02, 48.40, -129.70, -63.67) |
| Temperature-At-Depth Maps | Blackwell and others | Maps for the conterminous United States and Geothermal Resource Estimates; images of temperature at various depths. | (Blackwell and others, 2011) | Updates | Conterminous United States |
| Lithospheric Thickness from Sp Receiver Functions | Shallon | Lithospheric Thickness from Sp Receiver Functions | | In progress | (31.00, 43.00, -126.00, -112.00) |

The bounding box is (minimum latitude, maximum latitude, minimum longitude, maximum longitude) in degrees in the WGS84 horizontal datum.

CTM: Community Thermal Model

SCEC: Statewide California Earthquake Center

USGS: U.S. Geological Survey

Updates: Model updates are anticipated on a regular or irregular schedule.

In progress: Model is under initial development.

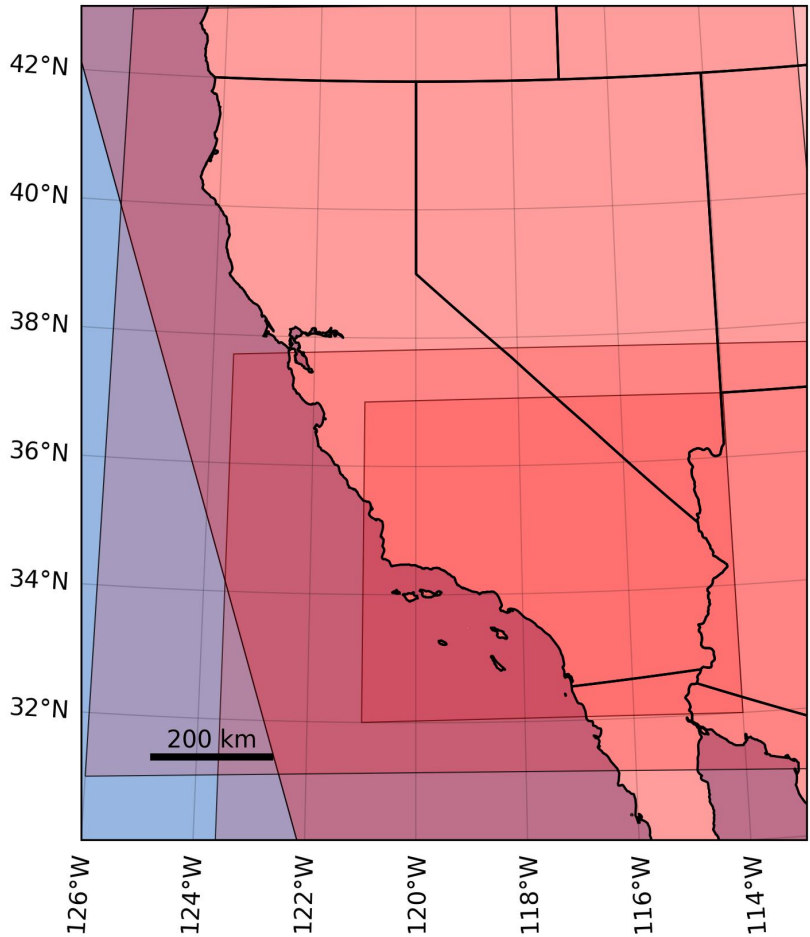

**Figure 5**: The shaded, semi-transparent polygons show the geographic coverage of the thermal models in Table 6.

## Stress models

Stress models for California (Figure 6 and Table 7) can be divided into three categories. Stressing rate models describe the stress accumulation rate on or near fault surfaces, informed by geodetic observations. Stress orientation models are mainly derived from the inversion of earthquake focal mechanisms and sometimes from borehole breakouts or other sources. Stress magnitude models are based on forward physics-based modeling of the tectonic loading and provide both magnitude and orientation estimates.

The SCEC Community Stress Model v2023 (Hardebeck and others, 2023) is a suite of models compiled over SCEC4 and SCEC5 (2012–2020) that span southern California. It includes five models of stressing rate and six models of stress, each with its own spatial extent, depth extent, assumptions, and limitations. All are available as a 3D Cartesian stress tensor and include useful derived metrics, such as principal stresses and orientations. There is also a curated set of SHmax (maximum horizontal stress) azimuths derived from borehole breakouts in the Los Angeles region (Luttrell and Hardebeck, 2021). Beyond this region, there are stress orientation models from focal mechanisms available for Ridgecrest (Hardebeck, 2020), the central and northern sections of the San Andreas Fault system (Provost and Houston, 2001; Provost and Houston, 2003), the greater San Francisco Bay region (Hardebeck and Michael, 2004), and the southeast San Francisco Bay region (Skoumal and others, 2023).

Stress orientation models are primarily consistent in regions of overlap. Stressing rate models agree near the main faults of the San Andreas Fault system, but there are substantial discrepancies away from these faults. The stress magnitude models disagree the most at depth, with estimates of differential stress varying by an order of magnitude.

Primary gaps in the coverage of stress models include sections of central, northern, and eastern California outside the boundaries of the original SCEC Community Stress Model consideration.

Some potential goals and milestones for community development of stress models in California include the following:

1. Prioritize filling in gaps in geographic coverage of the three types of stress-related models.
2. Evaluate the models in multiple applications and prioritize regions and features for improvement.
3. Assess alternative approaches for developing stress orientation, stress magnitude, and stressing rate models and integrating them with other types of community Earth models.

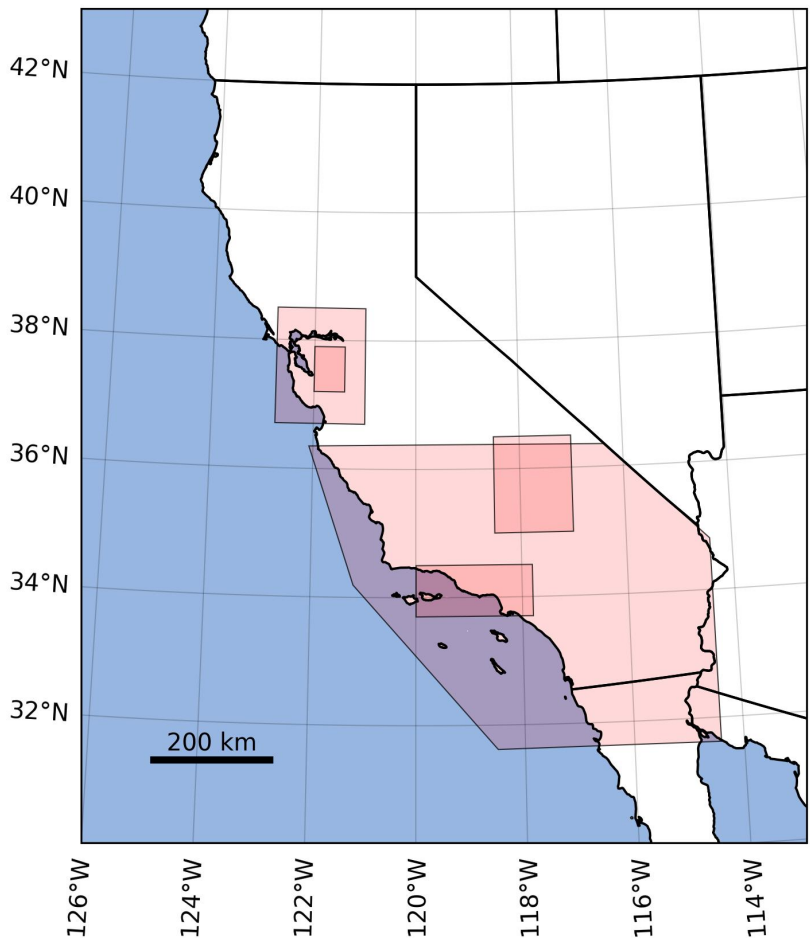

**Figure 6**: The shaded, semi-transparent polygons show the geographic coverage of the stress-related models in Table 7.

Table 7: Inventory of existing seismic stress models

| Name | Authors | Description | Reference | Status | Bounding Box |
|---|---|---|---|---|---|
| SCEC CSM | Hardebeck and others | Models of stress and stressing rate in southern California collected by SCEC CSM | (Hardebeck and others, 2023) | Updates | (31.62, 36.35, -122.08, -114.42) |
| LA Borehole Breakout SHmax | Luttrell and Hardebeck | SHmax orientations from LA area | (Luttrell and Hardebeck, 2021) | Published | (33.70, 34.50, -120.00, -117.80) |
| Ridgecrest stress orientation model | Hardebeck | Stress orientations inverted from earthquake focal mechanisms | (Hardebeck, 2020) | Published | (35.00, 36.50, -118.50, -117.00) |
| SF Bay region stress orientation model | Hardebeck and Michael | Stress orientations inverted from earthquake focal mechanisms | (Hardebeck and Michael, 2004) | Published | (36.70, 38.50, -122.75, -121.00) |
| SF Bay region stress orientation model | Skoumal and others | Stress orientations inverted from earthquake focal mechanisms | (Skoumal and others, 2023) | Published | (37.20, 37.90, -122.00, -121.40) |

The bounding box is (minimum latitude, maximum latitude, minimum longitude, maximum longitude) in degrees in the WGS84 horizontal datum.

CSM: Community Stress Model

LA: Los Angeles

SCEC: Statewide California Earthquake Center

SF: San Francisco

SHmax: Maximum horizontal stress

Updates: Model updates are anticipated on a regular or irregular schedule.

Published: Model is published, but no updates are anticipated.

# Geodetic models

Geodetic models for California from the late 1990s and early 2000s (Figure 7 and Table 8) were based primarily on Global Navigation Satellite Systems (GNSS) observations, with additional observations from electronic distance measurement or very long baseline interferometry data. These models include multiple generations of the SCEC Crustal Motion Map (CMM). They leveraged decades of survey GNSS data collection and data from an initially small but increasing number of continuous GNSS stations. These models provided estimates of average horizontal velocity at discrete locations across the region. The most recent version, CMM4 (Shen and others, 2011), included data from 1009 GNSS sites, spanning 1986–2004.

From the late 2000s onwards, synthetic aperture radar (SAR) data from satellites such as the European Remote-Sensing Satellites (ERS-1 and ERS-2), Envisat, and the Advanced Land Observing Satellite (ALOS) allowed estimates of surface deformation velocity using stacks of processed interferometric synthetic-aperture radar (InSAR) data. A series of studies ensued, such as the San Andreas Fault-wide study by Tong and others (2013), which used ALOS data from multiple tracks to estimate surface velocities on a dense grid of pixels. The velocity estimates produced this way were average surface velocities in the radar line-of-sight. Due to the significant uncertainties in the satellite orbits used in the InSAR processing, GNSS velocities were used to constrain the long-wavelength component of deformation and place the velocities estimated in an approximate reference frame.

Since 2014, geodetic models have developed in several directions. Routine GNSS processing from multiple processing centers has allowed models to expand geographic coverage to most of the western United States. For example, Zeng (2022) created a compilation of 4,979 continuous and campaign GNSS velocities supporting the 2023 National Seismic Hazard Model (Zeng, 2022). Developments in InSAR have produced more open-access and higher resolution SAR data than ever before, resulting in the creation of Sentinel-1 displacement products from Advanced Rapid Imaging and Analysis (https://aria.jpl.nasa.gov, accessed November 21, 2024) and Observational Products for End-Users from Remote Sensing Analysis (https://www.jpl.nasa.gov/go/opera/, accessed November 21, 2024), and a high-resolution displacement product from the Uninhabited Aerial Vehicle Synthetic Aperture Radar instrument (https://uavsar.jpl.nasa.gov/, accessed November 21, 2024).

The combination of GNSS and InSAR data has also rapidly advanced. For example, Shen and Liu (2020) produced a combination of GNSS and InSAR time series in southern California that constrained 3D deformation over four EnviSAT and ERS data tracks from 1992 to 2010. Xu and others (2021) produced a San Andreas Fault-wide Sentinel-1 velocity product from 2015 to 2019.5 (starting at the beginning of 2015 and ending midway through 2019) that utilized GNSS to constrain the long-wavelength deformation. Finally, the SCEC Community Geodetic Model version 2.0 has produced GNSS time series and InSAR time series and velocities over southern California from 2015 to 2019.5 in a consensus combination of multiple processing centers.

The past decade has also had more sophisticated geodetic models derived from simple velocity or time series products, such as interpolated strain rate fields and fault slip rate models. One example is the fault slip rates derived from a block modeling approach for the Eastern California Shear Zone (Hammond and others, 2024), although many others exist. The quintessential strain rate model used in the community is from Sandwell and others (2016), which is part of the SCEC Community Geodetic Model version 1. This version of the Community Geodetic Model covers southern California and explores the similarities and differences between 17 community-produced strain rate maps.

Future geodetic models face the joint challenges of geographic coverage, temporal coverage, and time-dependent deformation. For example, some models of southern California span only the period before the 2019 Ridgecrest earthquake, which was characterized by relatively simple linear motion, whereas others choose to constrain the complex spatial-temporal deformation pattern in the aftermath of the Ridgecrest mainshocks. Some models use geodetic data from earlier SAR satellites, such as EnviSAT and ERS, and others use the current generation, which includes Sentinel-1. Expanding regional models that include InSAR to statewide coverage would involve non-trivial computational costs. InSAR, based on the C-band satellites ERS, EnviSAT, and Sentinel-1, is inadequate for the densely vegetated parts of California, especially the northwest corner. The Japan Aerospace Exploration Agency

ALOS and ALOS-2 acquisitions only have good coverage from one look direction. The upcoming joint National Aeronautics and Space Administration and Indian Space Research Organization synthetic aperture radar (NISAR) mission will provide frequent coverage of the whole state from two look directions to enable more complete mapping of displacements.

Some potential goals and milestones for community development of geodetic models in California include the following:

1. Prioritize filling in gaps in the geographic coverage of models.
2. Identify discrepancies among models where they overlap, reconcile different models, and document alternative interpretations as appropriate.
3. Assess alternative approaches for removing transient effects in developing long-term deformation rates.

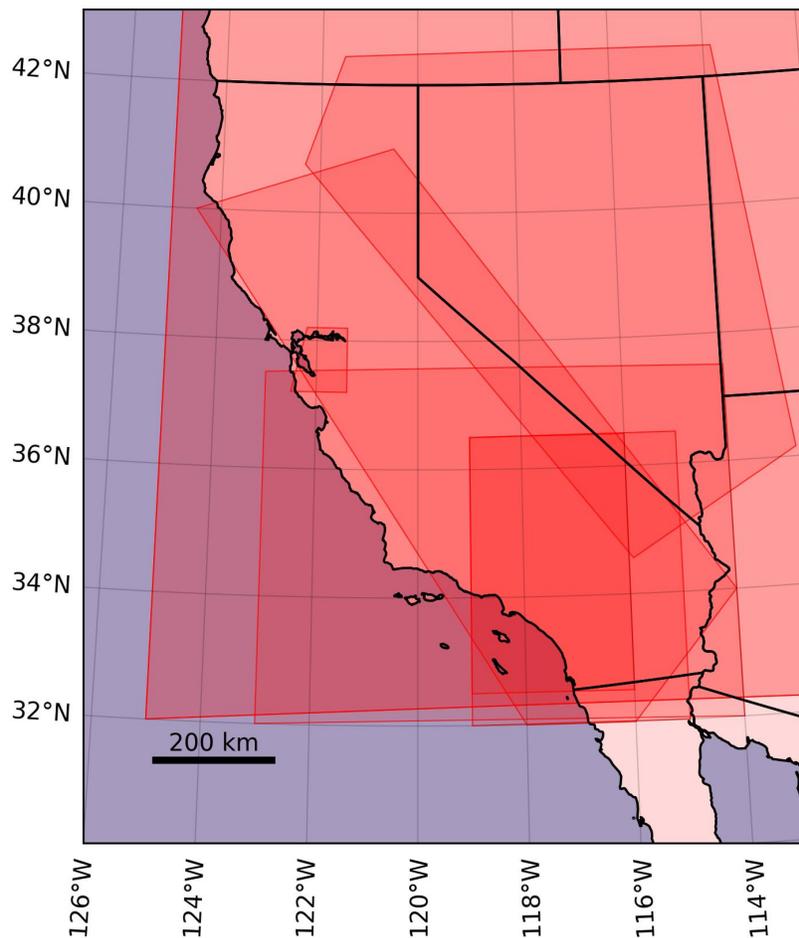

**Figure 7**: The shaded, semi-transparent polygons show the geographic coverage of the geodetic-related models in Table 8.

**Table 8**: Inventory of existing geodetic models.

| Name | Authors | Description | Reference | Status | Bounding Box |
|---|---|---|---|---|---|
| NSHM Velocity Field | Zeng | GNSS velocity field using 4979 continuous and campaign velocities from multiple sources | (Zeng, 2022) | Published | (28.38, 52.46, -128.60, -103.02) |
| USGS GNSS Velocities | Murray and Svarc | Continuous and campaign GNSS velocities processed by USGS | (Murray and Svarc, 2017) | Published | (32.00, 49.00, -125.00, -108.00) |
| ARIA and OPERA products | Jet Propulsion Laboratory team | Western United States covering operational Sentinel-1 processing | | In progress | (32.00, 49.00, -125.00, -108.00) |
| UAVSAR geodetic displacement dataset | Fielding and Zinke | High-resolution InSAR time series displacements from UAVSAR | | In progress | (37.20, 38.20, -122.50, -121.40) |
| Sentinel-1 velocity and time series product | Xu, Sandwell, Klein, Bock | Sentinel-1 velocity field integrated with GNSS velocity field for long wavelengths | (Xu and others, 2021) | Published | (32.00, 41.00, -124.50, -114.00) |
| InSAR and GNSS combined velocity 3D velocity | Shen and Liu | Four tracks of Envisat and ERS from 1992 to 2010 | (Shen and Liu, 2020) | Published | (32.50, 36.50, -119.0, -116.0) |
| SCEC CGM v2.0 | Floyd and others | GNSS time series and InSAR Sentinel-1 time series from 2015 to 2019.5 | Floyd and others, 2023 | Published | (32.00, 36.50, -119.0, -115.00) |
| Eastern California Slip Rates | Hammond, Kreemer, and Blewitt | Fault slip rate model from block modeling approach, constrained by GNSS velocities | (Hammond and others, 2024) | Published | (34.56, 42.49, -122.32, -112.69) |
| SCEC CGM v1.0 Strain rate comparison | Sandwell and others | Exercise to compare strain rate maps from various techniques | (Sandwell and others, 2016) | Published | (32.00, 37.50, -123.0, -114.0) |

The bounding box is (minimum latitude, maximum latitude, minimum longitude, maximum longitude) in degrees in the WGS84 horizontal datum.
3D: three dimensional
ARIA: Advanced Rapid Imaging and Analysis
CGM: Community Geodetic Model
ERS: European Remote-Sensing satellite
GNSS: Global Navigation Satellite Systems
InSAR: Interferometric Synthetic Aperture Rader

NSHM: National Seismic Hazard Model
OPERA: Observational Products for End-Users from Remote Sensing Analysis
SCEC: Statewide California Earthquake Center
UAVSAR: Uninhabited Aerial Vehicle Synthetic Aperture Radar
USGS: U.S. Geological Survey
Published: Model is published, but no updates are anticipated.
In progress: Model is under initial development.

## Techniques for integrating and embedding models

Model merging, embedding, and integration are essential for developing a suite of models that collectively provide seamless coverage for California. Many Earth models have been generated independently with different techniques, spatial extents, or resolutions. Often, the collocated values do not agree. Such discrepancies may arise from a need for more constraints or from different modeling techniques or assumptions.

Development of the SCEC Community Geodetic Model required addressing several of these issues. GNSS and InSAR data have different spatial and temporal resolutions and are merged to create the SCEC Community Geodetic Model. Methods proposed for aligning InSAR with GNSS (for example, Xu and others, 2021) could help GNSS inform the long wavelength part of the deformation field with shorter wavelength details from InSAR. Computation of the covariance between these datasets and tracking the uncertainties are critical. Other sources of deformation, such as hydrological loading, must also be removed.

Merging models with different spatial extents or embedding smaller, finer-resolution models into larger regional models is a common issue with the community seismic wave speed models. Values along the boundaries may not agree, creating artificial jumps in the elastic properties, leading to erroneous seismic wave propagation reflections. Similar issues likely will arise as new models are developed for the other types of community Earth models. Blending and tapering techniques can smooth these artificial jumps in elastic properties (Ajala and Persaud, 2021). A second approach leverages dictionary learning methods at different spatial scales to impose a geophysically consistent transition or incorporate higher-resolution features (Zhang and Ben-Zion, 2024a). A third approach involves building models using techniques that ensure consistency across models. For example, the detailed and regional domains of the USGS San Francisco Bay region seismic wave speed model (Aagaard and Hirakawa, 2021a; Hirakawa and Aagaard, 2022) are built on top of geologic models designed to have the same geologic units along the boundaries between the two models. This results in a seamless transition in elastic properties across the models.

## Connections with the Cascadia Region Earthquake Science Center

The transition of the Southern California Earthquake Center to the Statewide California Earthquake Center spanning the San Andreas Fault system and the new Cascadia Region Earthquake Science Center (CRESCENT) spanning the Cascadia subduction zone presents opportunities for integrating community Earth models along the west coast of the United States. The two centers share many objectives and use cases for community Earth models but have different challenges. California has more abundant seismicity, many damaging historical events, and a long history of data collection, whereas Cascadia has a lower rate of seismicity and emerging data coverage (including offshore). SCEC has developed community Earth models for over 30 years, whereas CRESCENT efforts are more nascent. CRESCENT's efforts may inspire fresh perspectives and workflows while benefitting from SCEC's experience addressing common obstacles.

The importance of findable, accessible, interoperable, and reproducible (FAIR) research practices (Wilkinson and others, 2016) is a priority as SCEC expands its geographic extent and CRESCENT builds its infrastructure. Striving for standard metadata and data formats for community Earth models developed by SCEC and CRESCENT would benefit both communities. There may be additional benefits in developing common infrastructure for exploring and merging overlapping community Earth models and sharing diverse use cases of community Earth models among the technical groups within each center to maximize the value of developing such models.

Both centers have an interest in consistent representation of the state of the crust and upper mantle around the Mendocino Triple Junction offshore northern California at about 40 degrees north. Collaboration between the centers presents an opportunity to leverage a wide range of expertise to advance our understanding of the behavior of the complex structure of this region with diverse faulting and seismic and aseismic slip (for example, Yeck and others, 2023; Yoon and Shelly, 2024).

## What does "community" in "community Earth models" mean?

Developing and maintaining community Earth models engages a broad spectrum of earth scientists, including researchers from private and academic sectors and professionals in earthquake-related industries. Belonging to the community advances science and opens participants to future and richer interdisciplinary collaborations. Community Earth models have been a central focus in SCEC for more than 30 years. They often differ from similar types of models produced by individual research groups in the following ways:

1. Models leverage multiple datasets generated by a variety of groups.
2. The community defines goals and milestones for improving the models.
3. Developers are motivated by the needs of the community.
4. Models are accessible and usable for a variety of applications.
5. Developers collaborate with the community to update models as methods are improved or new datasets become available.

Some groups within SCEC strive to build models, such as the SCEC Community Fault Model and SCEC Community Geodetic Model, that reflect the scientific consensus of the community. Other groups have focused on diverse representations of models, such as the SCEC Community Stress Model, to reflect epistemic uncertainty. Community Earth models ideally would document the workflow for contributors to submit new data or features for potential inclusion. Incorporating new data or features usually involves a series of discussions between the contributors and developers. In some cases, contradictory constraints or incompatible features may arise, and developers must make difficult choices to move forward. Such discussions may involve further analysis to (1) reconcile differences, (2) compromise to develop a consensus model, (3) create alternative models, or (4) postpone the incorporation of contributions until developers and contributors can resolve issues.

Developers of community Earth models often engage the community to evaluate models before release. Evaluation may involve developing community benchmarks, establishing general workflows for validation, or feedback from review panels or comments from the community.

All of these efforts rely on collaboration and benefit from engaging scientists from a variety of technical backgrounds and disciplines. Developing and maintaining *community* Earth models often requires more work and long-term commitment to their support and ongoing development compared with models developed for individual or small-group research projects.

## Incentives for participating in community Earth models

Community Earth models and open-source scientific software provide similar community benefits and share development attributes. In both cases, developers leverage their expertise to produce products that benefit a broad community. Over the past decade, methods for documenting contributions and attribution for models and software have become more standard. Nevertheless, developers and users of Earth models and scientific software would benefit from continuing cross-cutting discussions identifying best practices that address the community's needs for both types of products. In the rest of this section, we will focus on community Earth models, recognizing the discussion applies to both community Earth models and scientific software.

In the past, many researchers were both developers and users of community Earth models. It was common for graduate students to create an Earth model and apply it to address scientific questions. As the Earth models have become more sophisticated, scientists may focus solely on developing them. Making the Earth model publicly available with documentation provides powerful tools to others to advance science. Further development of an Earth model is generally contingent upon the authors' ability to document its use by others. Users can provide attribution in publications, which is most effective through Digital Object Identifiers (DOIs). Assigning DOIs requires archiving the Earth models in long-term repositories.

Many community Earth models rely on contributions from the community. Although researchers who are both developers and users often find immediate benefit to their contributions, providing adequate attribution for contributions is crucial, especially in cases where the contributors may not also be users. Recognizing contributors is an ongoing challenge. In some cases, contributors might be listed as coauthors on a publication or a citable data release. In other cases, contributions may not warrant being listed as coauthors, but recognition is still important. For example, authors may include a list of contributors in a data release. In all cases, using Open Researcher and Contributor ID (ORCiD) helps uniquely identify authors and contributors throughout their careers.

Regardless of how well contributions are documented, the success of community Earth models is driven by an open, collaborative, respectful scientific community dedicated to addressing common needs and advancing science.

## Outcomes and Recommendations

Community Earth models make decades of scientific expertise and knowledge accessible to a broad range of users and help drive cutting-edge breakthroughs. For example, a researcher modeling fault slip in any one of a variety of applications can leverage a fault geometry model that integrates geologic and geophysical data collected and analyzed over many decades. A scientist assessing local or regional seismic hazards may want to understand the geologic structure by examining geologic or seismic wave speed models.

### Accessibility

The most important indicator of the success of community Earth models is that the models are widely used to advance science or used in impactful scientific products such as seismic hazard models. Making models accessible to users with diverse technical backgrounds is critical and includes three main elements:

- Models are curated in public repositories, such as ScienceBase.gov and Zenodo.org, with appropriate versioning and Digital Object Identifiers.
- Documentation is provided for each model that includes
    - The workflow used to construct the model,
    - How to extract information from the model for common use cases,
    - How to contact developers and contributors with questions about the model, and
    - How members of the community can contribute to ongoing development.

  Model releases should include a list of contributors and their ORCiDs.

- Models should be disseminated using standard scientific data formats, such as Comma Separated Values, JSON, NetCDF (https://www.unidata.ucar.edu/software/netcdf/, accessed January 24, 2025), and HDF5 (https://www.hdfgroup.org/solutions/hdf5/, accessed January 24, 2025) appropriate for common use cases. Using standard scientific data formats allows users to incorporate the models easily into their workflows. Some users prefer a graphical interface to find and select models, whereas others prefer an application programming interface.

### Improving community Earth models

Several goals and milestones for improving community Earth models in California are shared across multiple types. These include the following:

- Extend the spatial coverage to span the entire state and relevant surrounding regions necessary to meet the needs of user applications.
- Identify discrepancies among models where they overlap and assess whether they represent epistemic uncertainty (viable alternatives that match observations) or whether the discrepancies can be resolved with constraints from existing observations.
- Develop techniques for quantifying the epistemic uncertainty and aleatory variability for the different types of models, leveraging common approaches, when possible, that facilitate the propagation of uncertainties in user applications.
- Prioritize geographic regions for improvement based on the epistemic uncertainty and seismic hazard and risk.

## Engaging the community

With the combined geographic footprint of SCEC and CRESCENT spanning the San Andreas Fault system and the Cascadia Subduction zone, there is an opportunity for the scientific community to produce a seamless suite of community Earth models spanning the west coast of the United States and extending well inland. Models are also developed by smaller working groups outside of these organizations. The scientific community benefits from cross-fertilization, data sharing, and collaboration among groups working on community Earth models. Some critical areas of cooperation could include the following:

- Leveraging common infrastructure for finding and accessing community Earth Models, such as graphical user interfaces, application programming interfaces, metadata, and data formats;
- Sharing data, techniques, and tools used to develop Earth models;
- Supporting users through online community forums and frequently asked questions; and
- Fostering community-driven development in which the developers engage a diverse group of stakeholders to establish priorities, milestones, and workflows for accepting contributions from the community.

Additionally, users benefit from regular briefings and discussions with developers on current work related to maintaining and updating community Earth models. These can take many forms, from discussions that are part of larger meetings to breakout or special sessions within larger meetings, workshops, or planned, informal gatherings at scientific conferences.

# Disclaimer

Any use of trade, firm, or product names is for descriptive purposes only and does not imply endorsement by the U.S. Government.

# References Cited

Aagaard, B.T., and Hirakawa, E.T., 2021a, San Francisco Bay region 3D seismic velocity model v21.0: U.S. Geological Survey data release, accessed February 28, 2024, at https://doi.org/10.5066/P98CA3D5.

Aagaard, B.T., and Hirakawa, E.T., 2021b, San Francisco Bay region 3D seismic velocity model v21.1: U.S. Geological Survey data release, accessed February 28, 2024, at https://doi.org/10.5066/P9TRDCHE.

Ahdi, S.K., Aagaard, B.T., Moschetti, M.P., Parker, G.A., Boyd, O.S., and Stephenson, W.J., 2024, Empirical response of select basins in the Western United States: Earthquake Spectra, v. 40, no. 2, p. 1099–1131, accessed April 15, 2024, at https://doi.org/10.1177/87552930241237250.

Ajala, R., and Persaud, P., 2021, Effect of merging multiscale models on seismic wavefield predictions near the southern San Andreas Fault: Journal of Geophysical Research: Solid Earth, v. 126, no. 10, p. e2021JB021915, accessed November 26, 2024, at https://doi.org/10.1029/2021JB021915.

Ben-Zion, Y., and Dresen, G., 2022, A synthesis of fracture, friction and damage processes in earthquake rupture zones: Pure and Applied Geophysics, v. 179, no. 12, p. 4323–4339, accessed January 31, 2025, at https://doi.org/10.1007/s00024-022-03168-9.

Blackwell, D., Richards, M., Frone, Z., Ruzo, A., Dingwall, R., and Williams, M., 2011, Temperature maps: accessed February 28, 2024, at Southern Methodist University Dedman College of Humanities and Sciences at https://www.smu.edu/dedman/academics/departments/earth-sciences/research/geothermallab/datamaps/temperaturemaps.

Bowden, D.C., Kohler, M.D., Tsai, V.C., and Weeraratne, D.S., 2016, Offshore Southern California lithospheric velocity structure from noise cross-correlation functions: Journal of Geophysical Research: Solid Earth, v. 121, no. 5, p. 3415–3427, accessed August 15, 2024, at https://doi.org/10.1002/2016JB012919.

Boyd, O.S., 2019, Grids in support of the U.S. Geological Survey thermal model for seismic hazard studies: U.S. Geological Survey data release, accessed February 28, 2024, at https://doi.org/10.5066/P935DT1G.

Boyd, O.S., and Shah, A.K., 2018, Components of the USGS National Crustal Model: U.S. Geological Survey data release, accessed December 15, 2022, at https://doi.org/10.5066/P9T96Q67.

Bozorgnia, Y., Abrahamson, N.A., Ahdi, S.K., Ancheta, T.D., Atik, L.A., Archuleta, R.J., Atkinson, G.M., Boore, D.M., Campbell, K.W., S-J Chiou, B., Contreras, V., Darragh, R.B., Derakhshan, S., Donahue, J.L., and others, 2022, NGA-Subduction research program: Earthquake Spectra, v. 38, no. 2, p. 783–798, accessed December 13, 2024, at https://doi.org/10.1177/87552930211056081.

Bozorgnia, Y., Abrahamson, N.A., Atik, L.A., Ancheta, T.D., Atkinson, G.M., Baker, J.W., Baltay, A., Boore, D.M., Campbell, K.W., Chiou, B.S.-J., Darragh, R., Day, S., Donahue, J., Graves, R.W., and others, 2014, NGA-West2 research project: Earthquake Spectra, v. 30, no. 3, p. 973–987, accessed December 14, 2022, at https://doi.org/10.1193/072113EQS209M.

Bradley, B., Bora, S., Lee, R., Manea, E.F., Gerstenberger, M.C., Stafford, P.J., Atkinson, G.M., Weatherill, G., Hutchinson, J., and de la Torre, C., 2022, Summary of the ground-motion characterisation model for the 2022 New Zealand National Seismic Hazard Model: GNS Science report 2022/46, 44 pp., accessed February 12, 2025, at https://doi.org/10.21420/9BMK-ZK64.

Burov, E.B., 2011, Rheology and strength of the lithosphere: Marine and Petroleum Geology, v. 28, no. 8, p. 1402–1443, accessed January 31, 2025, at https://doi.org/10.1016/j.marpetgeo.2011.05.008.

Castellanos, J.C., 2019, Dataset S1 - P-wave velocity model of Long Beach, CA: CaltechDATA [data set], accessed February 12, 2025, at https://doi.org/10.22002/D1.1293.

Castellanos, J.C., 2021, S-wave velocity model of Long Beach, CA: CaltechDATA [data set], accessed February 12, 2025, at https://doi.org/10.22002/D1.1970.

Catchings, R.D., Goldman, M.R., Li, Y. -G., and Chan, J.H., 2016, Continuity of the West Napa–Franklin fault zone inferred from guided waves generated by earthquakes following the 24 August 2014 Mw 6.0 South Napa earthquake: Bulletin of the Seismological Society of America, v. 106, no. 6, p. 2721–2746, accessed January 27, 2025, at https://doi.org/10.1785/0120160154.

Cheng, C.H., and Johnston, D.H., 1981, Dynamic and static moduli: Geophysical Research Letters, v. 8, no. 1, p. 39–42, accessed January 31, 2025, at https://doi.org/10.1029/GL008i001p00039.

Choi, J.-H., Edwards, P., Ko, K., and Kim, Y.-S., 2016, Definition and classification of fault damage zones: A review and a new methodological approach: Earth-Science Reviews, v. 152, p. 70–87, accessed January 31, 2025, at https://doi.org/10.1016/j.earscirev.2015.11.006.

Cromwell, G., Sweetkind, D.S., Langenheim, V.E., and Ely, C.P., 2024, Three-dimensional hydrogeologic framework model of the Russian River watershed, California: U.S. Geological Survey Scientific Investigations Report 2024–5083, 25 pp., accessed January 29, 2025, at https://doi.org/10.3133/sir20245083.

Dawson, T.E., 2013, Updates to the California Reference Fault Parameter Database—Uniform California Earthquake Rupture Forecast, Version 3 Fault Models 3.1 and 3.2, in Field, E.H. and others (eds.) Uniform California earthquake rupture forecast, version 3 (UCERF3): The time-independent model: U.S. Geological Survey Open-File Report 2013–1165 Appendix A, 18 pp., accessed February 28, 2024, at https://pubs.usgs.gov/of/2013/1165/pdf/ofr2013-1165_appendixA.pdf.

Doody, C., 2023, Dataset for "CANVAS: An adjoint waveform tomography model of California and Nevada": Zenodo, accessed February 28, 2024, at https://doi.org/10.5281/zenodo.8415562.

Doody, C., Rodgers, A., Afanasiev, M., Boehm, C., Krischer, L., Chiang, A., and Simmons, N., 2023, CANVAS: An Adjoint Waveform Tomography Model of California and Nevada: Journal of Geophysical Research: Solid Earth, v. 128, no. 12, p. e2023JB027583, accessed August 15, 2024, at https://doi.org/10.1029/2023JB027583.

Evans, W.S., Plesch, A., Shaw, J.H., Pillai, N.L., Yu, E., Meier, M., and Hauksson, E., 2020, A statistical method for associating earthquakes with their source faults in Southern California: Bulletin of the Seismological Society of America, v. 110, no. 1, p. 213–225, accessed, August 15, 2024, at https://doi.org/10.1785/0120190115.

Field, E.H., Arrowsmith, R.J., Biasi, G.P., Bird, P., Dawson, T.E., Felzer, K.R., Jackson, D.D., Johnson, K.M., Jordan, T.H., Madden, C., Michael, A.J., Milner, K.R., Page, M.T., Parsons, T., and others, 2014, Uniform California Earthquake Rupture Forecast (ver. 3) (UCERF3)—The time-independent model: Bulletin of the Seismological Society of America, v. 104, no. 3, p. 1122–1180, accessed January 27, 2023, at https://doi.org/10.1785/0120130164.

Field, E.H., Milner, K.R., Hatem, A.E., Powers, P.M., Pollitz, F.F., Llenos, A.L., Zeng, Y., Johnson, K.M., Shaw, B.E., McPhillips, D., Thompson Jobe, J., Shumway, A.M., Michael, A.J., Shen, Z., and others, 2023, The USGS 2023 conterminous U.S. time-independent earthquake rupture forecast: Bulletin of the Seismological Society of America, v. 114, no. 1, p. 523–571, accessed February 28, 2024, at https://doi.org/10.1785/0120230120.

Gabriel, A., Ulrich, T., Marchandon, M., Biemiller, J., and Rekoske, J., 2023, 3D dynamic rupture modeling of the 6 February 2023, Kahramanmaraş, Turkey Mw 7.8 and 7.7 earthquake doublet using early observations: The Seismic Record, v. 3, no. 4, p. 342–356, accessed November 20, 2024, at https://doi.org/10.1785/0320230028.

Gerstenberger, M.C., Bora, S., Bradley, B.A., DiCaprio, C., Kaiser, A., Manea, E.F., Nicol, A., Rollins, C., Stirling, M.W., Thingbaijam, K.K.S., Van Dissen, R.J., Abbott, E.R., Atkinson, G.M., Chamberlain, C., and others, 2023, The 2022 Aotearoa New Zealand National Seismic Hazard Model: Process, overview, and results: Bulletin of the Seismological Society of America, v. 114, no. 1, p. 7–36, accessed January 27, 2025, at https://doi.org/10.1785/0120230182.

Gooch, B., 2022, 3D geological modeling: California Department of Conservation, accessed February 28, 2024, at https://www.conservation.ca.gov/cgs/3DGeo.

Graves, R.W., and Aagaard, B.T., 2011, Testing long-period ground-motion simulations of scenario earthquakes using the Mw 7.2 El Mayor–Cucapah mainshock: Evaluation of finite-fault rupture characterization and 3D seismic velocity models: Bulletin of the Seismological Society of America, v. 101, no. 2, p. 895–907, accessed February 28, 2024, at https://doi.org/10.1785/0120100233.

Graves, R., Jordan, T.H., Callaghan, S., Deelman, E., Field, E., Juve, G., Kesselman, C., Maechling, P., Mehta, G., Milner, K., Okaya, D., Small, P., and Vahi, K., 2010, CyberShake: A physics-based seismic hazard model for southern California: Pure and Applied Geophysics, v. 168, no. 3, p. 367–381, accessed January 27, 2025, at https://doi.org/10.1007/s00024-010-0161-6.

Hammond, W.C., Kreemer, C., and Blewitt, G., 2024, Robust imaging of fault slip rates in the Walker Lane and Western Great Basin from GPS data using a multi-block model approach: Journal of Geophysical Research: Solid Earth, v. 129, no. 3, p. e2023JB028044, accessed November 21, 2024, at https://doi.org/10.1029/2023JB028044.

Hardebeck, J.L., 2020, A stress-similarity triggering model for aftershocks of the Mw 6.4 and 7.1 Ridgecrest earthquakes: Bulletin of the Seismological Society of America, v. 110, no. 4, p. 1716–1727, accessed August 15, 2024, at https://doi.org/10.1785/0120200015.

Hardebeck, J.L., Becker, T., Bird, P., Cooke, M., Hauksson, E., Hearn, E., Johnson, K., Loveless, J., Luttrell, K., Meade, B., Shen, Z., Smith-Konter, B., Yang, W., and Zeng, Y., 2023, SCEC Community Stress Model (CSM): Zenodo, accessed February 28, 2024, at https://zenodo.org/records/8270631.

Hardebeck, J.L., and Michael, A.J., 2004, Stress orientations at intermediate angles to the San Andreas Fault, California: Journal of Geophysical Research: Solid Earth, v. 109, no. B11, p. B11303, accessed August 15, 2024, at https://doi.org/10.1029/2004JB003239.

Hatem, A.E., Collett, C.M., Briggs, R.W., Gold, R.D., Angster, S.J., Powers, P.M., Field, E.H., Anderson, M., Ben-Horin, J.Y., Dawson, T., DeLong, S., DuRoss, C., Thompson Jobe, J., Kleber, E., and others, 2023, Earthquake geology inputs for the U.S. National Seismic Hazard Model (NSHM) 2023 (western U.S.) (ver. 3.0, December 2023): U.S. Geological Survey data release, accessed February 28, 2024, at https://doi.org/10.5066/P9AWINWZ.

Hearn, E., Montesi, L., Oskin, M., Hirth, G., Thatcher, W., and Behr, W., 2020, SCEC Community Rheology Model (CRM): Zenodo, accessed February 28, 2024, at https://doi.org/10.5281/zenodo.4579626.

Hearn, E., Oskin, M., Montesi, L., Hirth, G., Behr, W., Thatcher, W., and Legg, M., 2022, SCEC Community Rheology Model (CRM): Statewide California Earthquake Center, accessed February 28, 2024, at https://southern.scec.org/research/crm.

Hirakawa, E., and Aagaard, B., 2022, Evaluation and updates for the USGS San Francisco Bay region 3D seismic velocity model in the east and north bay portions: Bulletin of the Seismological Society of America, v. 112, no. 4, p. 2070–2096, accessed December 1, 2023, at https://doi.org/10.1785/0120210256.

Jachens, R.C., 2006, 3-D geologic and seismic velocity models of the San Francisco Bay region: U.S. Geological Survey Earthquake Hazards Program, accessed February 28, 2024, at https://www.usgs.gov/programs/earthquake-hazards/science/3-d-geologic-and-seismic-velocity-models-san-francisco-bay.

Jacquey, A.B., and Cacace, M., 2020, Multiphysics modeling of a brittle-ductile lithosphere: 2. Semi-brittle, semi-ductile deformation and damage rheology: Journal of Geophysical Research: Solid Earth, v. 125, no. 1, p. e2019JB018475, accessed January 31, 2025, at https://doi.org/10.1029/2019JB018475.

Jaeger, J.C., Cook, N.G.W., and Zimmerman, R., 2007, Fundamentals of Rock Mechanics, 4th Edition: Wiley-Blackwell, 488 pp.

Jia, Z., 2020, LAS1 wave velocity model for the central Los Angeles Basin: CaltechDATA [data set], accessed February 12, 2025, at https://doi.org/10.22002/D1.1670.

Jia, Z., Jin, Z., Marchandon, M., Ulrich, T., Gabriel, A.-A., Fan, W., Shearer, P., Zou, X., Rekoske, J., Bulut, F., Garagon, A., and Fialko, Y., 2023, The complex dynamics of the 2023 Kahramanmaraş, Turkey, Mw 7.8-7.7 earthquake doublet: Science, v. 381, no. 6661, p. 985–990, accessed November 20, 2024, at https://doi.org/10.1126/science.adi0685.

Jordan, T.H., Callaghan, S., Graves, R.W., Wang, F., Milner, K.R., Goulet, C.A., Maechling, P.J., Olsen, K.B., Cui, Y., Juve, G., Vahi, K., Yu, J., Deelman, E., and Gill, D., 2018, Cybershake models of seismic hazards in southern and central California, *in* Proceedings of the 11th National Conference in Earthquake Engineering: Earthquake Engineering Research Institute, Los Angeles, CA.

Karrech, A., Regenauer-Lieb, K., and Poulet, T., 2011, Continuum damage mechanics for the lithosphere: Journal of Geophysical Research: Solid Earth, v. 116, no. B4, p. B04205, accessed January 31, 2025, at https://doi.org/10.1029/2010JB007501.

Kim, A., Dreger, D.S., and Larsen, S., 2010, Moderate earthquake ground-motion validation in the San Francisco Bay area: Bulletin of the Seismological Society of America, v. 100, no. 2, p. 819–825, accessed January 27, 2025, at https://doi.org/10.1785/0120090076.

Kim, Y.-S., Peacock, D.C.P., and Sanderson, D.J., 2004, Fault damage zones: Journal of Structural Geology, v. 26, no. 3, p. 503–517, accessed January 31, 2025, at https://doi.org/10.1016/j.jsg.2003.08.002.

Lee, E., Chen, P., and Jordan, T.H., 2014, Testing waveform predictions of 3D velocity models against two recent Los Angeles earthquakes: Seismological Research Letters, v. 85, no. 6, p. 1275–1284, accessed January 27, 2025, at https://doi.org/10.1785/0220140093.

Li, Y.-G., Aki, K., Vidale, J.E., Lee, W.H.K., and Marone, C.J., 1994, Fine structure of the Landers fault zone: segmentation and the rupture process: Science, v. 265, no. 5170, p. 367–370, accessed January 27, 2025, at https://doi.org/10.1126/science.265.5170.367.

Li, G., and Ben-Zion, Y., 2024, Multi-scale velocity models for the Ridgecrest region, CA: Zenodo, accessed August 15, 2024, at https://doi.org/10.5281/zenodo.10740714.

Luttrell, K., and Hardebeck, J., 2021, A unified model of crustal stress heterogeneity from borehole breakouts and earthquake focal mechanisms: Journal of Geophysical Research: Solid Earth, v. 126, no. 2, p. e2020JB020817, accessed August 15, 2024, at https://doi.org/10.1029/2020JB020817.

Magistrale, H., McLaughlin, K., and Day, S., 1996, A geology-based 3-D velocity model of the Los Angeles basin sediments: Bulletin of the Seismological Society of America, v. 86, no. 4, p. 1161–1166, accessed February 28, 2025, at https://doi.org/10.1785/BSSA0860041161.

Manaker, D.M., Turcotte, D.L., and Kellogg, L.H., 2006, Flexure with damage: Geophysical Journal International, v. 166, no. 3, p. 1368–1383, accessed January 31, 2025, at https://doi.org/10.1111/j.1365-246X.2006.03067.x.

Meletti, C., Marzocchi, W., D'Amico, V., Lanzano, G., Luzi, L., Martinelli, F., Pace, B., Rovida, A., Taroni, M., and Visini, F., 2021, The new Italian seismic hazard model (MPS19): Annals of Geophysics, v. 64, no. 1, p. SE112, accessed January 27, 2025, at https://doi.org/10.4401/ag-8579.

Morrison, F.A., 2001, Understanding Rheology: Oxford University Press, 545 pp.

Moschetti, M.P., Aagaard, B.T., Ahdi, S.K., Altekruse, J., Boyd, O.S., Frankel, A.D., Herrick, J., Petersen, M.D., Powers, P.M., Rezaeian, S., Shumway, A.M., Smith, J.A., Stephenson, W.J., Thompson, E.M., and others, 2024, The 2023 US National Seismic Hazard Model: Ground-motion characterization for the conterminous United States: Earthquake Spectra, v. 40, no. 2, p. 1158–1190, accessed March 3, 2025, at https://doi.org/10.1177/87552930231223995.

Muir, J.B., Clayton, R.W., Tsai, V.C., and Brissaud, Q., 2021, Tomographic model for the northeastern Los Angeles Basin: CaltechDATA [data set], accessed February 12, 2025, at https://doi.org/10.22002/D1.2088.

Murray, J.R., and Svarc, J., 2017, Global positioning system data collection, processing, and analysis conducted by the U.S. Geological Survey Earthquake Hazards Program: Seismological Research Letters, v. 88, no. 3, p. 916–925, accessed November 21, 2024, at https://doi.org/10.1785/0220160204.

Olsen, K.B., and Mayhew, J.E., 2010, Goodness-of-fit criteria for broadband synthetic seismograms, with application to the 2008 Mw 5.4 Chino Hills, California, earthquake: Seismological Research Letters, v. 81, no. 5, p. 715–723, accessed January 27, 2025, at https://doi.org/10.1785/gssrl.81.5.715.

Oskin, M., 2024, Geological Framework Model Viewer (Provisional): Statewide California Earthquake Center, accessed February 28, 2024, at http://moho.scec.org/GFM_web/web/viewer.php.

Pan, Y.W., and Wen, B.H., 2015, Constitutive model for the continuous damage of brittle rock: Géotechnique, v. 51, no. 2, p. 155–159, accessed January 31, 2025, at https://doi.org/10.1680/geot.2001.51.2.155.

Persaud, P., 2022, 3-D basement, Z2.5 and P-wave velocity model of Coachella and Imperial valleys, Southern California: University of Arizona Department of Geosciences, accessed February 28, 2024, at https://www.geo.arizona.edu/~ppersaud/Data.html.

Petersen, M.D., Shumway, A.M., Powers, P.M., Field, E.H., Moschetti, M.P., Jaiswal, K.S., Frankel, A.D., Rezaeian, S., Llenos, A.L., Michael, A., Boyd, O.S., Rukstales, K.S., Clayton, B.S., Altekruse, J.M., and others, 2024, The 2023 U.S. 50-state National Seismic Hazard Model: Overview and implications:

Earthquake Spectra, v. 40, no. 1, p. 5–88, accessed January 31, 2025, at https://doi.org/10.1177/87552930231215428.

Plesch, A., Marshall, S., and Shaw, J., 2023, The SCEC Community Fault Model (CFM): Statewide California Earthquake Center, accessed February 28, 2024, at https://southern.scec.org/research/cfm.

Plesch, A., Shaw, J.H., Benson, C., Bryant, W.A., Carena, S., Cooke, M., Dolan, J., Fuis, G., Gath, E., Grant, L., Hauksson, E., Jordan, T., Kamerling, M., Legg, M., and others, 2007, Community Fault Model (CFM) for Southern California: Bulletin of the Seismological Society of America, v. 97, no. 6, p. 1793–1802, accessed August 15, 2024, at https://doi.org/10.1785/0120050211.

Plesch, A., Shaw, J.H., Ross, Z.E., and Hauksson, E., 2020, Detailed 3D fault representations for the 2019 Ridgecrest, California, earthquake sequence: Bulletin of the Seismological Society of America, v. 110, no. 4, p. 1818–1831, accessed November 20, 2024, at https://doi.org/10.1785/0120200053.

Provost, A.-S., and Houston, H., 2001, Orientation of the stress field surrounding the creeping section of the San Andreas Fault: Evidence for a narrow mechanically weak fault zone: Journal of Geophysical Research: Solid Earth, v. 106, no. B6, p. 11373–11386, accessed August 16, 2024, at https://doi.org/10.1029/2001JB900007.

Provost, A.-S., and Houston, H., 2003, Stress orientations in northern and central California: Evidence for the evolution of frictional strength along the San Andreas plate boundary system: Journal of Geophysical Research: Solid Earth, v. 108, no. B3, p. 2175, accessed August 16, 2024, at https://doi.org/10.1029/2001JB001123.

Rekoske, J.M., Gabriel, A.-A., and May, D.A., 2023, Instantaneous physics-based ground motion maps using reduced-order modeling: Journal of Geophysical Research: Solid Earth, v. 128, no. 8, p. e2023JB026975, accessed November 20, 2024, at https://doi.org/10.1029/2023JB026975.

Riesner, M., Durand-Riard, P., Hubbard, J., Plesch, A., and Shaw, J.H., 2017, Building objective 3D fault representations in active tectonic settings: Seismological Research Letters, v. 88, no. 3, p. 831–839, accessed November 20, 2024, at https://doi.org/10.1785/0220160192.

Rutter, E.H., and Brodie, K.H., 1991, Lithosphere rheology—a note of caution: Journal of Structural Geology, v. 13, no. 3, p. 363–367, accessed January 31, 2025, at https://doi.org/10.1016/0191-8141(91)90136-7.

Sandwell, D.T., Zeng, Y., Shen, Z., Crowell, B.W., Murray, J.R., McCaffrey, R., and Xu, X., 2016, The SCEC Community Geodetic Model V1: Horizontal velocity grid: 2016 Southern California Earthquake Center Annual Meeting, Palm Springs, California, accessed February 28, 2024, at https://central.scec.org/publication/6967.

Savage, J.C., 1990, Equivalent strike-slip earthquake cycles in half-space and lithosphere-asthenosphere earth models: Journal of Geophysical Research: Solid Earth, v. 95, no. B4, p. 4873–4879, accessed January 31, 2025, at https://doi.org/10.1029/JB095iB04p04873.

SB-905 Carbon sequestration: Carbon Capture, Removal, Utilization, and Storage Program, 2022: California Legislative Information, accessed November 20, 2024, at https://leginfo.legislature.ca.gov/faces/billNavClient.xhtml?bill_id=202120220SB905.

Scheirer, A.H., 2007, The three-dimensional geologic model used for the 2003 National Oil and Gas Assessment of the San Joaquin Basin Province, California, in Scheirer, A.H. (ed.) Petroleum Systems and Geologic Assessment of Oil and Gas in the San Joaquin Basin Province, California: U.S. Geological Survey Professional Paper 1713, Chapter 7, 81 pp., accessed January 25, 2025, at https://doi.org/10.3133/pp1713.

Schliwa, N., and Gabriel, A., 2023, Equivalent near-field corner frequency analysis of 3D dynamic rupture simulations reveals dynamic source effects: Seismological Research Letters, v. 95, no. 2A, p. 900–924, accessed November 20, 2024, at https://doi.org/10.1785/0220230225.

Seebeck, H., Van Dissen, R., Litchfield, N., Barnes, P.M., Nicol, A., Langridge, R., Barrell, D.J.A., Villamor, P., Ellis, S., Rattenbury, M., Bannister, S., Gerstenberger, M., Ghisetti, F., Sutherland, R., and others, 2024, The New Zealand Community Fault Model – version 1.0: An improved geological foundation for seismic hazard modelling: New Zealand Journal of Geology and Geophysics, v. 67, no. 2, p. 209–229, accessed August 15, 2024, at https://doi.org/10.1080/00288306.2023.2181362.

Shen, S., Gao, Y., and Jia, L., 2024, A comparison of the relationship between dynamic and static rock mechanical parameters: Applied Sciences, v. 14, no. 11, p. 4487, accessed January 31, 2025, at https://doi.org/10.3390/app14114487.

Shen, Z.-K., King, R.W., Agnew, D.C., Wang, M., Herring, T.A., Dong, D., and Fang, P., 2011, A unified analysis of crustal motion in Southern California, 1970–2004: The SCEC crustal motion map: Journal of Geophysical Research: Solid Earth, v. 116, no. B11, p. B11402, accessed November 21, 2024, at https://doi.org/10.1029/2011JB008549.

Shen, Z.-K., and Liu, Z., 2020, Integration of GPS and InSAR data for resolving 3-dimensional crustal deformation: Earth and Space Science, v. 7, no. 4, p. e2019EA001036, accessed November 21, 2024, at https://doi.org/10.1029/2019EA001036.

Shinevar, W.J., Behn, M.D., Hirth, G., and Jagoutz, O., 2018, Inferring crustal viscosity from seismic velocity: Application to the lower crust of Southern California: Earth and Planetary Science Letters, v. 494, p. 83–91, accessed August 15, 2024, at https://doi.org/10.1016/j.epsl.2018.04.055.

Simmons, G., and Brace, W.F., 1965, Comparison of static and dynamic measurements of compressibility of rocks: Journal of Geophysical Research (1896-1977), v. 70, no. 22, p. 5649–5656, accessed January 31, 2025, at https://doi.org/10.1029/JZ070i022p05649.

Skoumal, R.J., Hardebeck, J.L., and Shelly, D.R., 2023, Using corrected and imputed polarity measurements to improve focal mechanisms in a regional earthquake catalog near the Mt. Lewis fault zone, California: Journal of Geophysical Research: Solid Earth, v. 128, no. 2, p. e2022JB025660, accessed August 15, 2024, at https://doi.org/10.1029/2022JB025660.

Small, P., Gill, D., Maechling, P.J., Taborda, R., Callaghan, S., Jordan, T.H., Olsen, K.B., Ely, G.P., and Goulet, C., 2017, The SCEC Unified Community Velocity Model software framework: Seismological Research Letters, v. 88, no. 6, p. 1539–1552, accessed November 26, 2024, at https://doi.org/10.1785/0220170082.

Southern California Earthquake Center (SCEC), 2021, Southern California Velocity Model developed by the Harvard Structural Geology Group (SCEC CVM-H): Statewide California Earthquake Center, accessed February 28, 2024, at https://github.com/SCECcode/cvmh/.

Southern California Earthquake Center (SCEC), 2022a, Southern California Earthquake Center Central California Seismic Velocity Model (SCEC CCA06): Statewide California Earthquake Center, accessed February 28, 2024, at https://github.com/SCECcode/CCA.

Southern California Earthquake Center (SCEC), 2022b, The Southern California Velocity Model with geotechnical layer (SCEC CVM-S4): Statewide California Earthquake Center, accessed February 28, 2024, at https://github.com/SCECcode/cvms.

Southern California Earthquake Center (SCEC), 2022c, The tomography improved Southern California Velocity Model with geotechnical layer (cvmsi): Statewide California Earthquake Center, accessed February 28, 2024, at https://github.com/SCECcode/cvmsi.

Spudich, P., and Olsen, K.B., 2001, Fault zone amplified waves as a possible seismic hazard along the Calaveras Fault in central California: Geophysical Research Letters, v. 28, no. 13, p. 2533–2536, accessed January 27, 2025, at https://doi.org/10.1029/2000GL011902.

Sweetkind, D.S., and Faunt, C.C., 2024, Digital data for 3-D geologic framework and textural models for Cuyama Valley groundwater basin, California: U.S. Geological Survey data release, accessed January 29, 2025, at https://doi.org/10.5066/P1NAABFG.

Sweetkind, D.S., and Zellman, K.L., 2022, Spatial data from an inventory of U.S. Geological Survey three-dimensional geologic models, Volume 1, 2004–2022: U.S. Geological Survey data release, accessed February 28, 2024, at https://doi.org/10.5066/P9OFRNUN.

Taborda, R., and Bielak, J., 2014, Ground-motion simulation and validation of the 2008 Chino Hills, California, earthquake using different velocity models: Bulletin of the Seismological Society of America, v. 104, no. 4, p. 1876–1898, accessed January 27, 2025, at https://doi.org/10.1785/0120130266.

Taufiqurrahman, T., Gabriel, A.-A., Li, D., Ulrich, T., Li, B., Carena, S., Verdecchia, A., and Gallovič, F., 2023, Dynamics, interactions and delays of the 2019 Ridgecrest rupture sequence: Nature, v. 618, no. 7964, p. 308–315, accessed November 20, 2024, at https://doi.org/10.1038/s41586-023-05985-x.

Thatcher, W., Chapman, D., Hearn, E., and Shinevar, W.J., n.d., SCEC Community Thermal Model (CTM): Statewide California Earthquake Center, accessed February 28, 2024, at https://southern.scec.org/research/ctm.

Thatcher, W., Shinevar, W.J., Chapman, D., and Hearn, E., 2020, SCEC Community Thermal Model (CTM): Zenodo, accessed February 28, 2024, at https://doi.org/10.5281/zenodo.4010833.

Tong, X., Sandwell, D.T., and Smith-Konter, B., 2013, High-resolution interseismic velocity data along the San Andreas Fault from GPS and InSAR: Journal of Geophysical Research: Solid Earth, v. 118, no. 1, p. 369–389, accessed November 21, 2024, at https://doi.org/10.1029/2012JB009442.

Traum, J.A., Teague, N.F., Sweetkind, D.S., and Nishikawa, T., 2022, Hydrologic and geochemical characterization of the Petaluma River watershed, Sonoma County, California: U.S. Geological Survey Scientific Investigations Report 2022-5009, 217 pp., accessed January 29, 2025, at https://doi.org/10.3133/sir20225009.

U.S. Geological Survey, 2022, Quaternary Fault and Fold Database of the United States: U.S. Geological Survey

Earthquake Hazards Program, accessed February 28, 2024, at https://www.usgs.gov/programs/earthquake-hazards/faults.

Walton, M.A.L., Papesh, A.G., Johnson, S.Y., Conrad, J.E., and Brothers, D.S., 2020, Quaternary faults offshore of California: U.S. Geological Survey data release, accessed February 28, 2024, at https://doi.org/10.5066/P91RYEZ4.

Wilkinson, M.D., Dumontier, M., Aalbersberg, Ij.J., Appleton, G., Axton, M., Baak, A., Blomberg, N., Boiten, J.-W., da Silva Santos, L.B., Bourne, P.E., Bouwman, J., Brookes, A.J., Clark, T., Crosas, M., and others, 2016, The FAIR guiding principles for scientific data management and stewardship: Scientific Data, v. 3, no. 1, p. 160018, accessed January 20, 2023, at https://doi.org/10.1038/sdata.2016.18.

Xu, X., Sandwell, D.T., Klein, E., and Bock, Y., 2021, Integrated Sentinel-1 InSAR and GNSS time-series along the San Andreas Fault system: Journal of Geophysical Research: Solid Earth, v. 126, no. 11, p. e2021JB022579, accessed November 21, 2024, at https://doi.org/10.1029/2021JB022579.

Yeck, W.L., Shelly, D.R., Materna, K.Z., Goldberg, D.E., and Earle, P.S., 2023, Dense geophysical observations reveal a triggered, concurrent multi-fault rupture at the Mendocino Triple Junction: Communications Earth & Environment, v. 4, p. 94, accessed November 26, 2024, at https://doi.org/10.1038/s43247-023-00752-2.

Yoon, C.E., and Shelly, D.R., 2024, Distinct yet adjacent earthquake sequences near the Mendocino Triple Junction: 20 December 2021 Mw 6.1 and 6.0 Petrolia, and 20 December 2022 Mw 6.4 Ferndale: The Seismic Record, v. 4, no. 1, p. 81–92, accessed November 26, 2024, at https://doi.org/10.1785/0320230053.

Zeng, Y., 2022, GPS velocity field of the western United States for the 2023 National Seismic Hazard Model update: Seismological Research Letters, v. 93, no. 6, p. 3121–3134, accessed November 21, 2024, at https://doi.org/10.1785/0220220180.

Zhang, H., and Ben-Zion, Y., 2024a, Enhancing regional seismic velocity models with higher-resolution local results using sparse dictionary learning: Journal of Geophysical Research: Solid Earth, v. 129, no. 1, p. e2023JB027016, accessed November 26, 2024, at https://doi.org/10.1029/2023JB027016.

Zhang, H., and Ben-Zion, Y., 2024b, Merged multiscale seismic velocity models for southern California (CVM-S4.26-Fang_etal_2019): Zenodo, accessed August 15, 2024, at https://doi.org/10.5281/zenodo.10778491.

# Appendix A: Geographic extent of models in inventory

## Geologic models

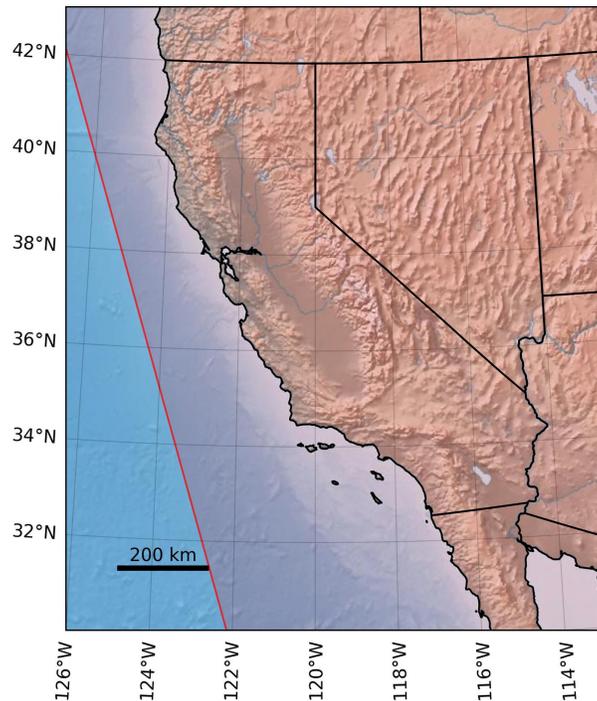

**Figure A1**: The shaded, semi-transparent polygon shows the geographic coverage of the geologic model for the U.S. Geological Survey (USGS) National Crustal Model listed in Table 2. Background image from Natural Earth (naturalearth.com).

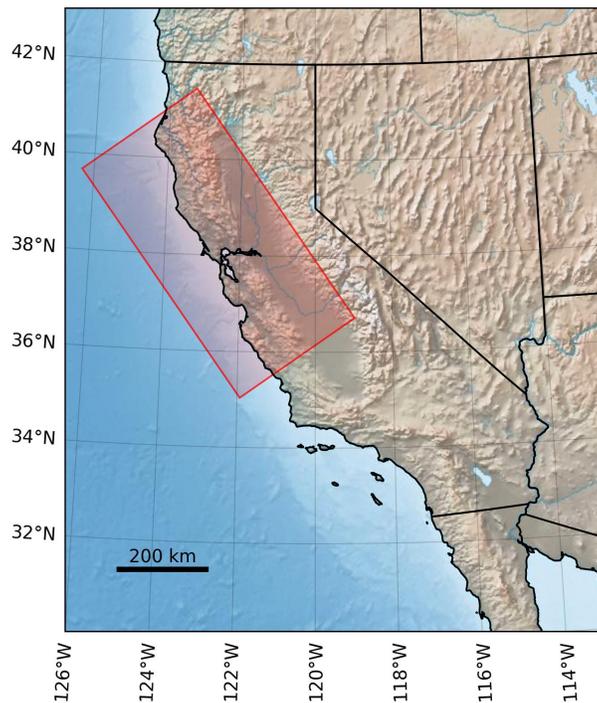

**Figure A2**: The shaded, semi-transparent polygon shows the geographic coverage of the geologic model for the U.S. Geological Survey (USGS) San Francisco Bay regional domain listed in Table 2. Background image from Natural

Earth (naturalearth.com).

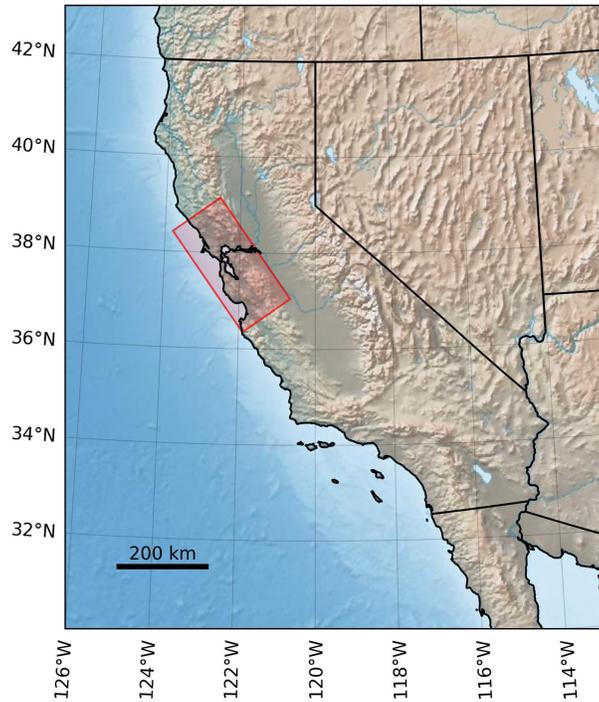

**Figure A3**: The shaded, semi-transparent polygon shows the geographic coverage of the geologic model for the U.S. Geological Survey (USGS) San Francisco Bay detailed domain listed in Table 2. Background image from Natural Earth (naturalearth.com).

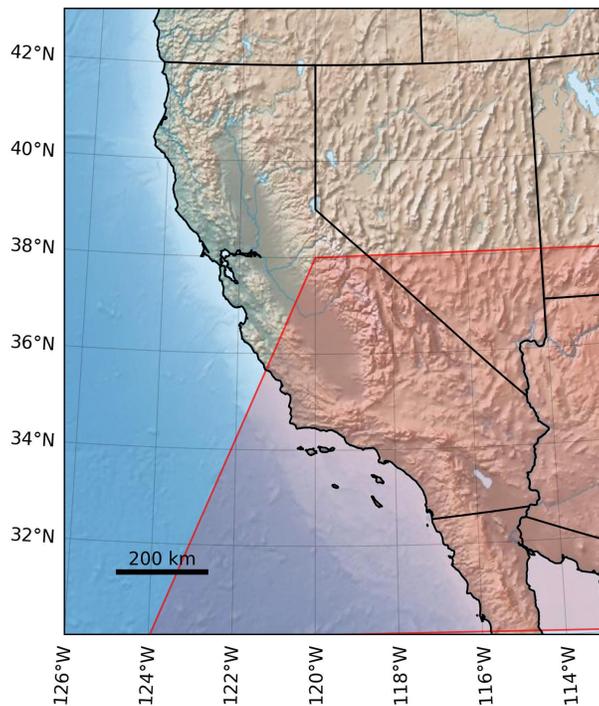

**Figure A4**: The shaded, semi-transparent polygon shows the geographic coverage of the Statewide California Earthquake Center (SCEC) Geologic Framework Model listed in Table 2. Background image from Natural Earth (naturalearth.com).

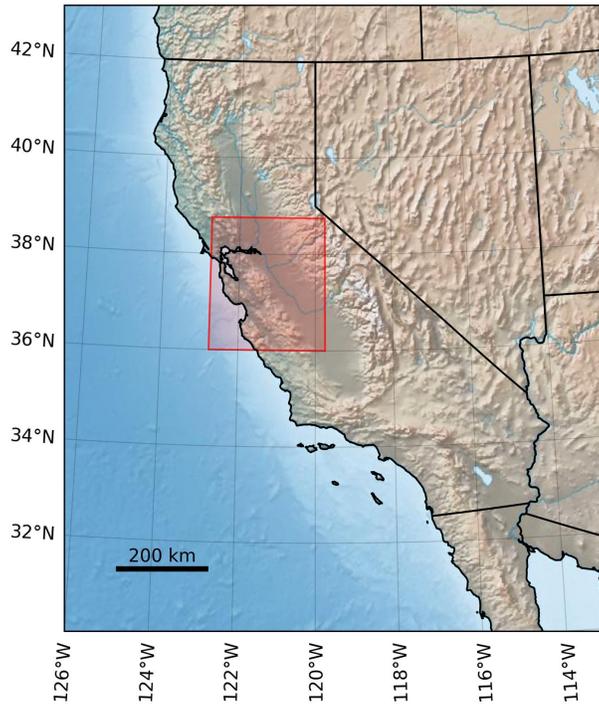

**Figure A5**: The shaded, semi-transparent polygon shows the geographic coverage of the Medwedeff San Francisco geologic model listed in Table 2. Background image from Natural Earth (naturalearth.com).

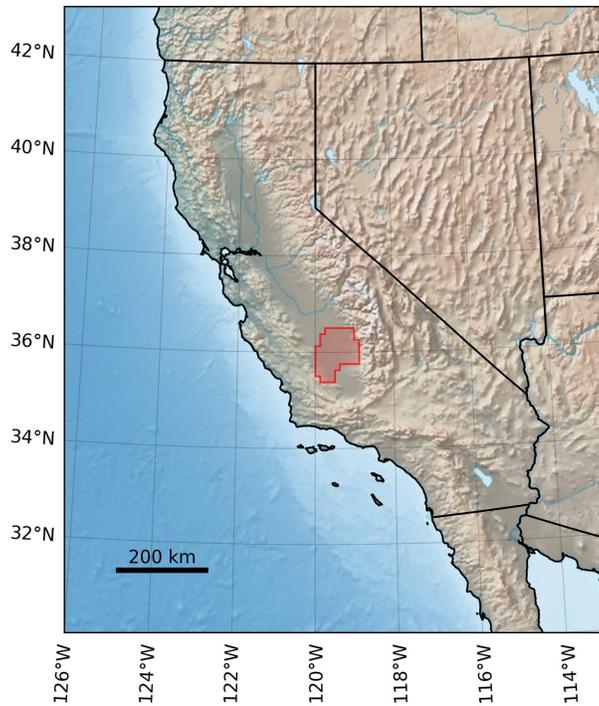

**Figure A6**: The shaded, semi-transparent polygon shows the geographic coverage of the southern San Joaquin Geologic Framework Model listed in Table 2. Background image from Natural Earth (naturalearth.com).

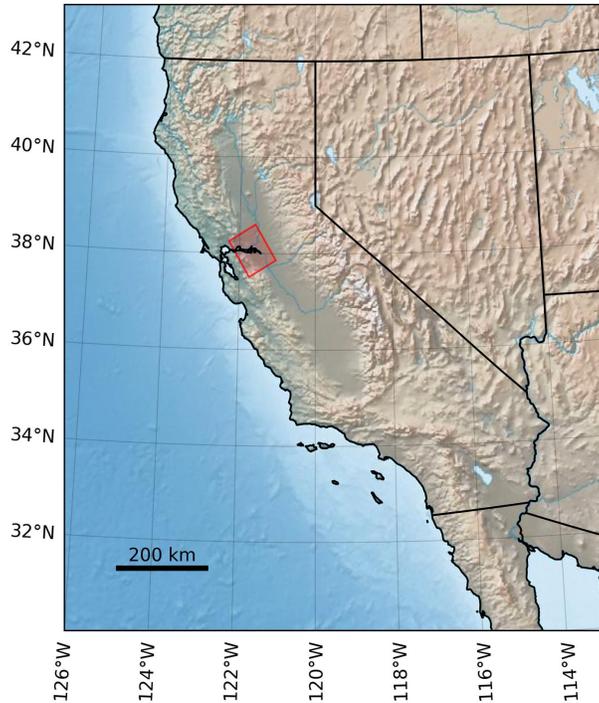

**Figure A7**: The shaded, semi-transparent polygon shows the geographic coverage of the geologic model of the Sacramento-San Joaquin River delta listed in Table 2. Background image from Natural Earth (naturalearth.com).

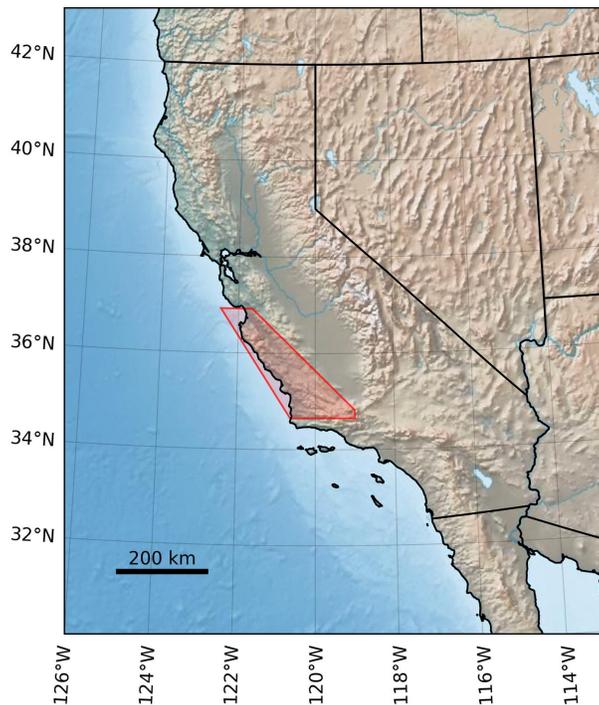

**Figure A8**: The shaded, semi-transparent polygon shows the geographic coverage of the geologic model of the Central Coast Ranges listed in Table 2. Background image from Natural Earth (naturalearth.com).

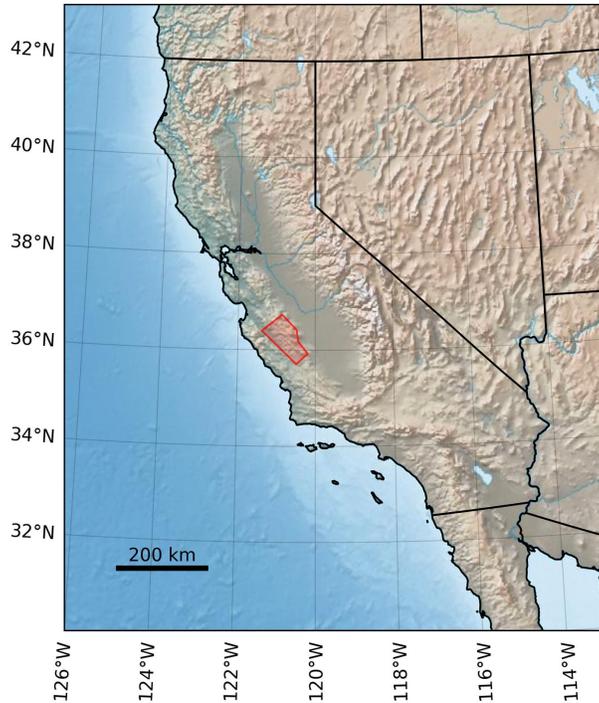

**Figure A9**: The shaded, semi-transparent polygon shows the geographic coverage of the geologic model of the San Andreas Fault Zone listed in Table 2. Background image from Natural Earth (naturalearth.com).

# Fault geometry models

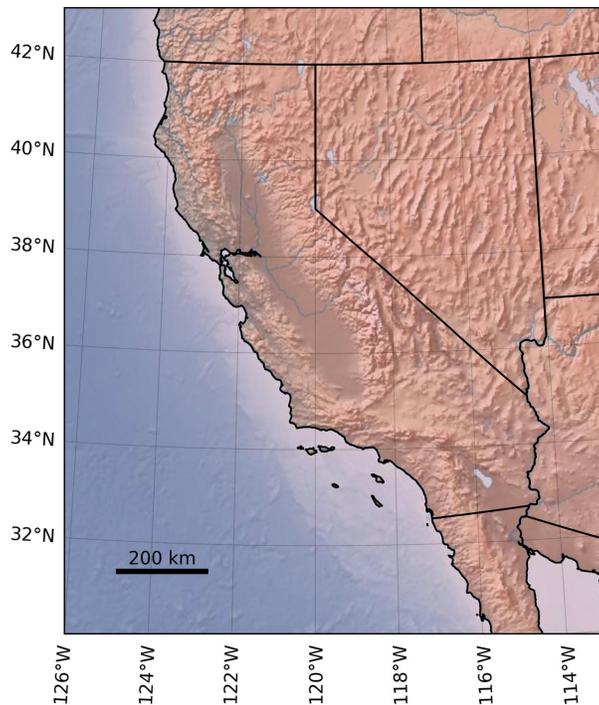

**Figure A10**: The shaded, semi-transparent polygon (which covers the entire map) shows the geographic coverage of the fault geometry for the 2023 National Seismic Hazard Model (NSHM23) listed in Table 3. Background image from Natural Earth (naturalearth.com).

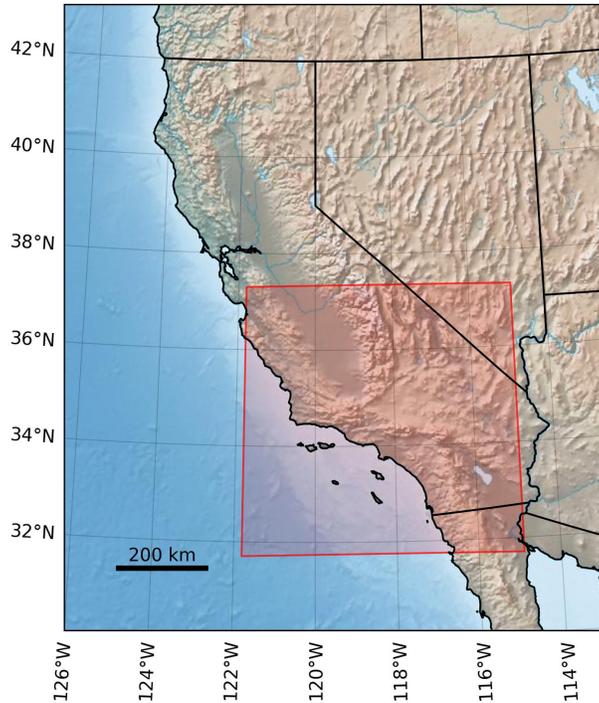

**Figure A11**: The shaded, semi-transparent polygon shows the geographic coverage of the Statewide California Earthquake Center (SCEC) Community Fault Model listed in Table 3. Background image from Natural Earth (naturalearth.com).

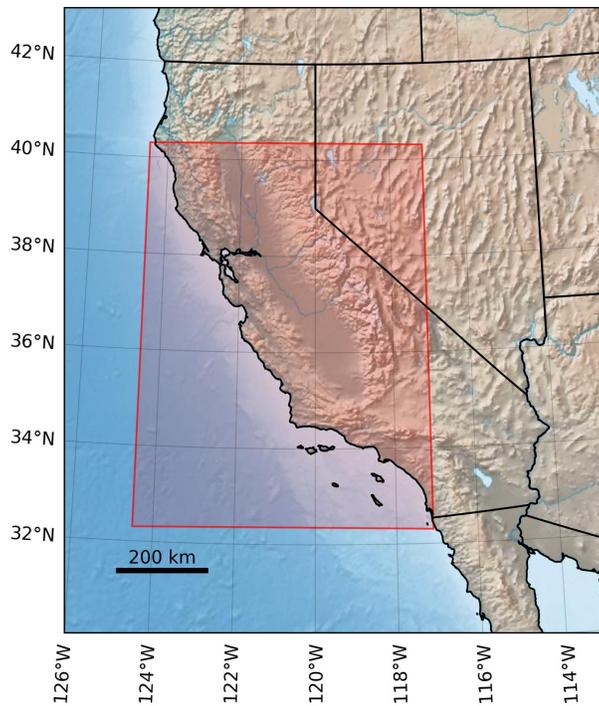

**Figure A12**: The shaded, semi-transparent polygon shows the geographic coverage of the fault geometry model Quaternary faults offshore of California listed in Table 3. Background image from Natural Earth (naturalearth.com).

## Seismic wave speed models

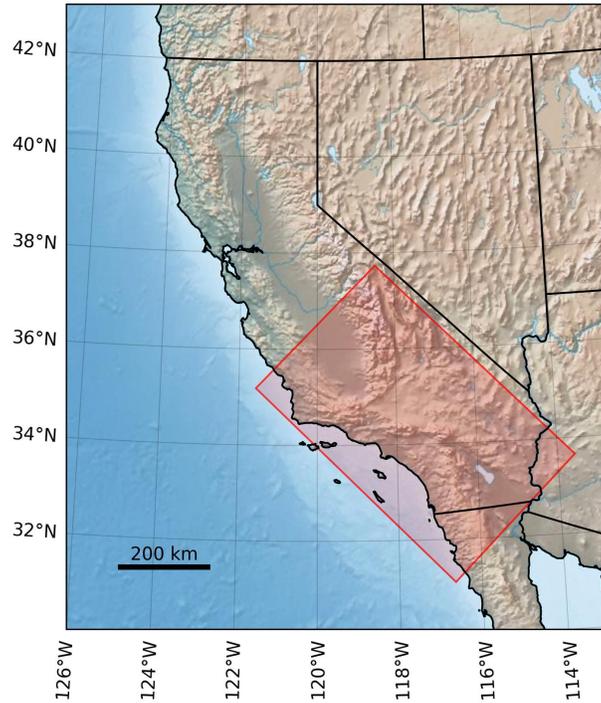

**Figure A13**: The shaded, semi-transparent polygon shows the geographic coverage of the Statewide California Earthquake Center (SCEC) Community Velocity Model (CVM) S4 listed in Table 4. Background image from Natural Earth (naturalearth.com).

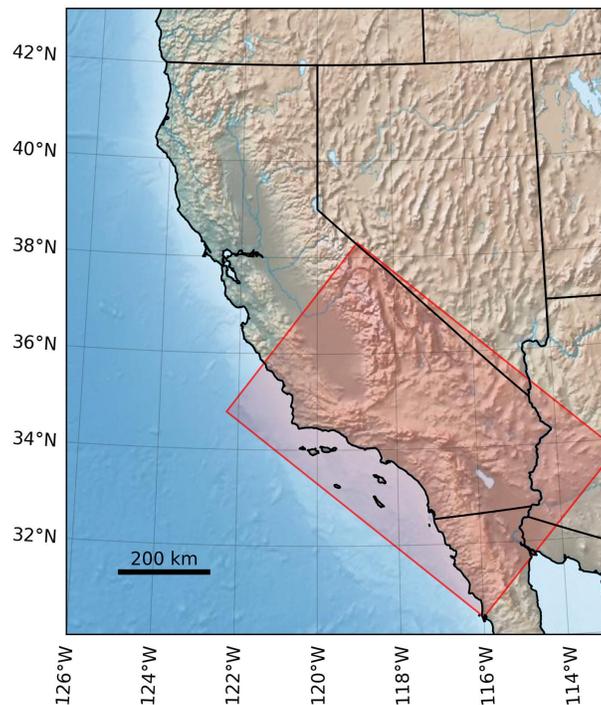

**Figure A14**: The shaded, semi-transparent polygon shows the geographic coverage of the Statewide California Earthquake Center (SCEC) Community Velocity Model (CVM) S4.26.M01 listed in Table 4. Background image from Natural Earth (naturalearth.com).

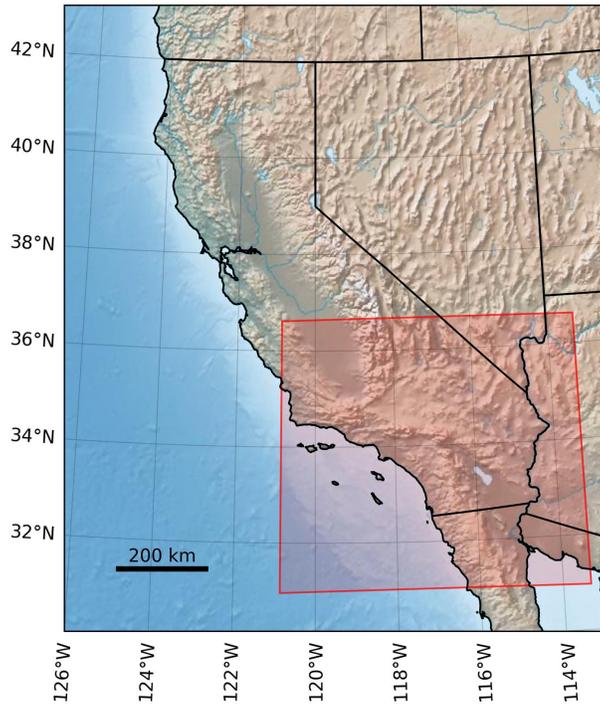

**Figure A15**: The shaded, semi-transparent polygon shows the geographic coverage of the Statewide California Earthquake Center (SCEC) Community Velocity Model (CVM) H listed in Table 4. Background image from Natural Earth (naturalearth.com).

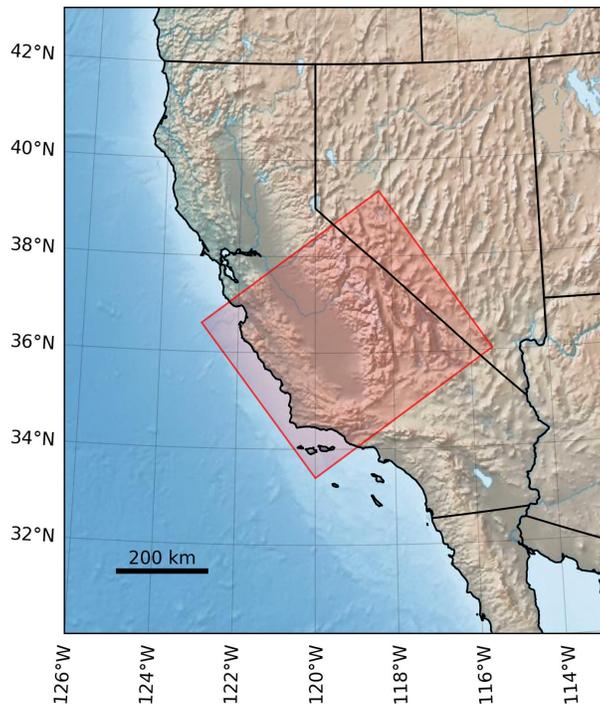

**Figure A16**: The shaded, semi-transparent polygon shows the geographic coverage of the Statewide California Earthquake Center (SCEC) Central California (CCA) 06 seismic wavespeed model listed in Table 4. Background image from Natural Earth (naturalearth.com).

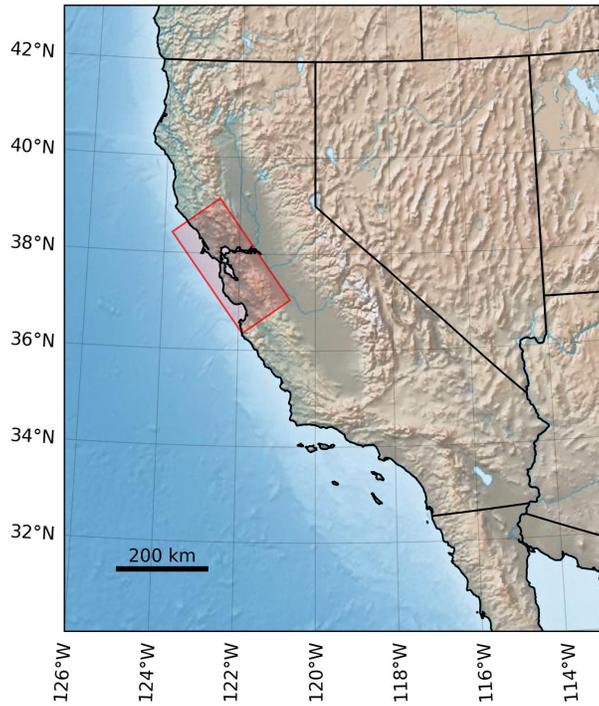

**Figure A17**: The shaded, semi-transparent polygon shows the geographic coverage of the seismic wave speed model for the USGS San Francisco Bay detailed domain listed in Table 4. Background image from Natural Earth (naturalearth.com).

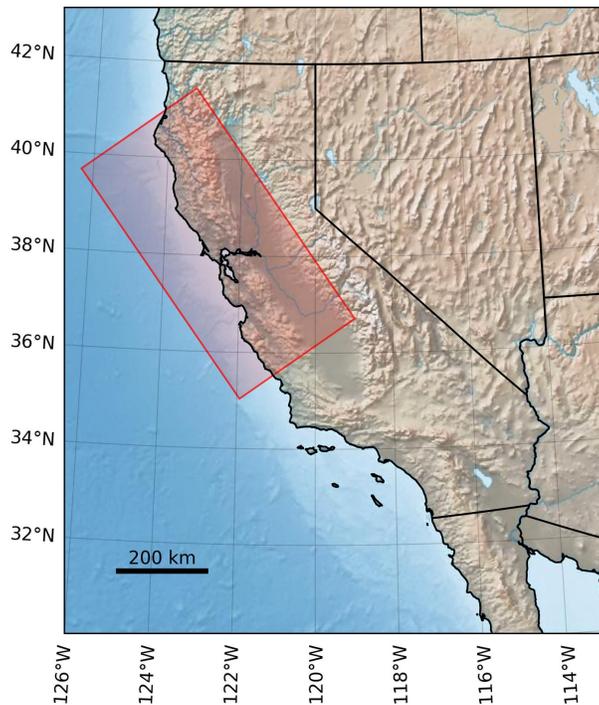

**Figure A18**: The shaded, semi-transparent polygon shows the geographic coverage of the seismic wave speed model for the U.S. Geological Survey (USGS) San Francisco Bay regional domain listed in Table 4. Background image from Natural Earth (naturalearth.com).

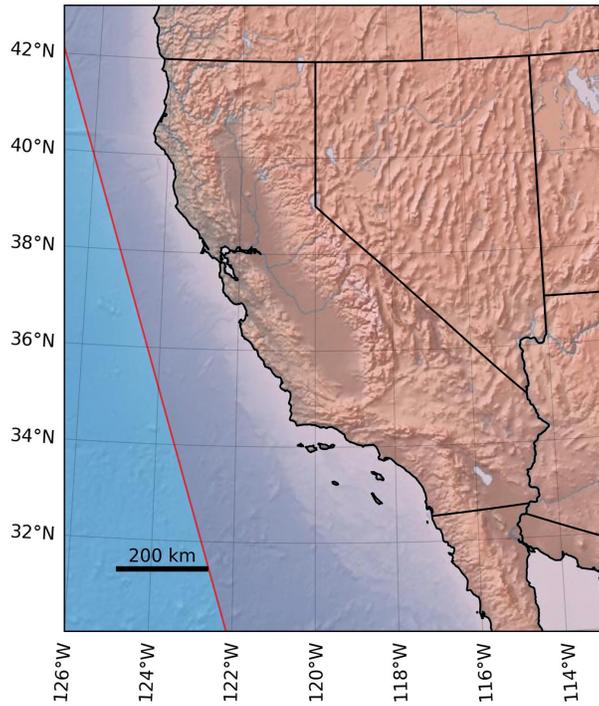

**Figure A19**: The shaded, semi-transparent polygon shows the geographic coverage of the seismic wave speed model for the U.S. Geological Survey (USGS) National Crustal Model (NCM) listed in Table 4. Background image from Natural Earth (naturalearth.com).

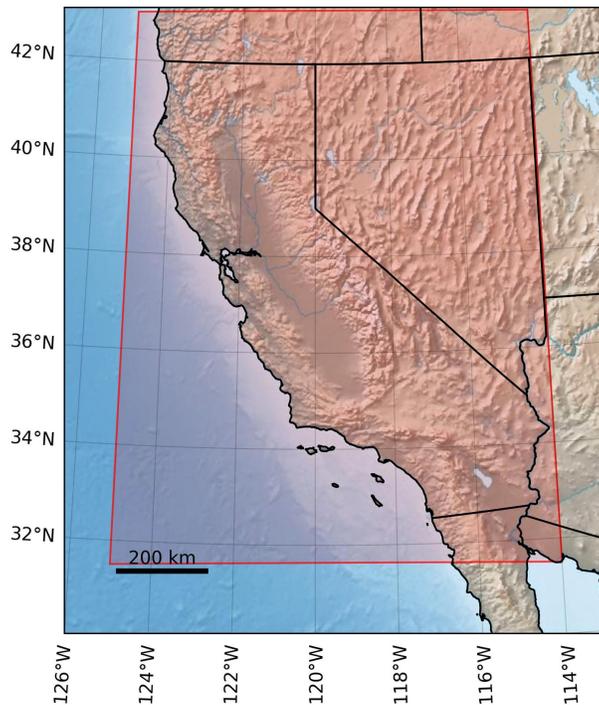

**Figure A20**: The shaded, semi-transparent polygon shows the geographic coverage of the CANVAS seismic wave speed model listed in Table 4. Background image from Natural Earth (naturalearth.com).

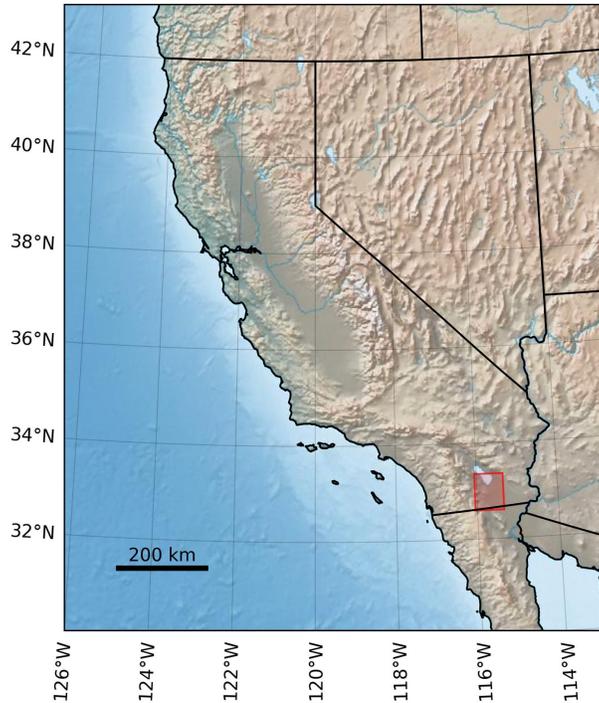

**Figure A21**: The shaded, semi-transparent polygon shows the geographic coverage of the Salton Sea Imaging Project (SSIP) Imperial seismic wave speed model listed in Table 4. Background image from Natural Earth (naturalearth.com).

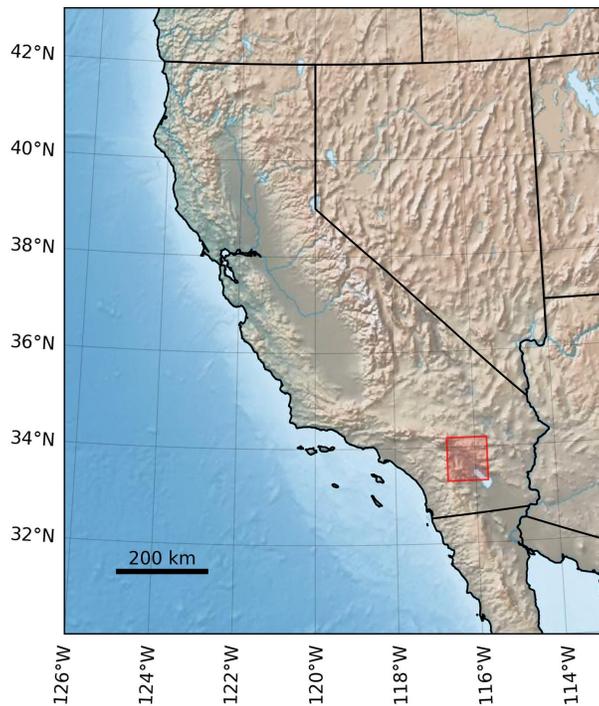

**Figure A22**: The shaded, semi-transparent polygon shows the geographic coverage of the Salton Sea Imaging Project (SSIP) Coachella seismic wave speed model listed in Table 4. Background image from Natural Earth (naturalearth.com).

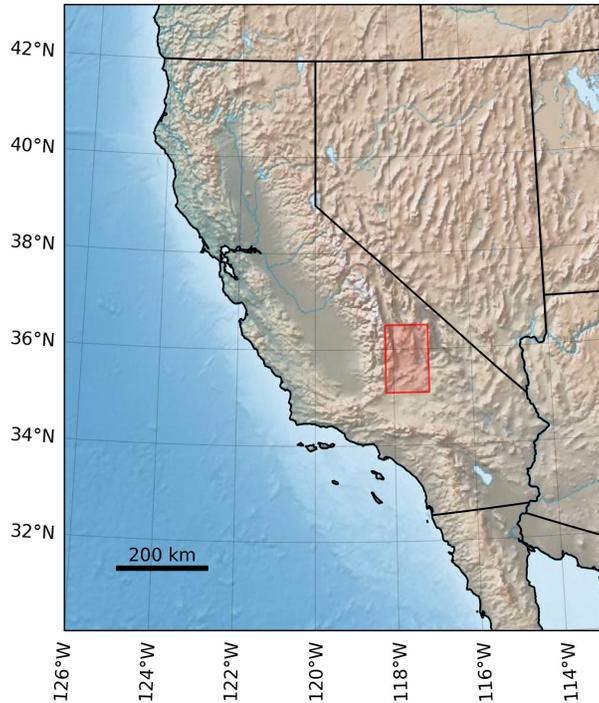

**Figure A23**: The shaded, semi-transparent polygon shows the geographic coverage of the Li and Ben-Zion seismic wave speed model listed in Table 4. Background image from Natural Earth (naturalearth.com).

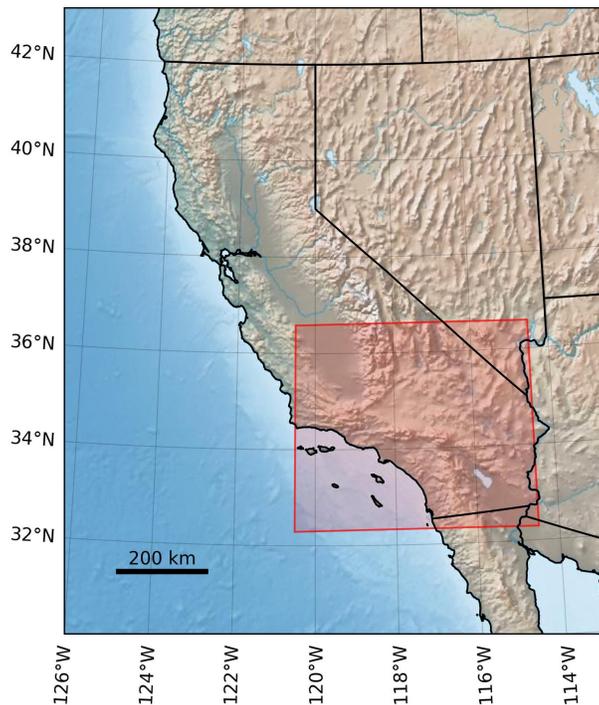

**Figure A24**: The shaded, semi-transparent polygon shows the geographic coverage of the Zhang and Ben-Zion seismic wave speed model listed in Table 4. Background image from Natural Earth (naturalearth.com).

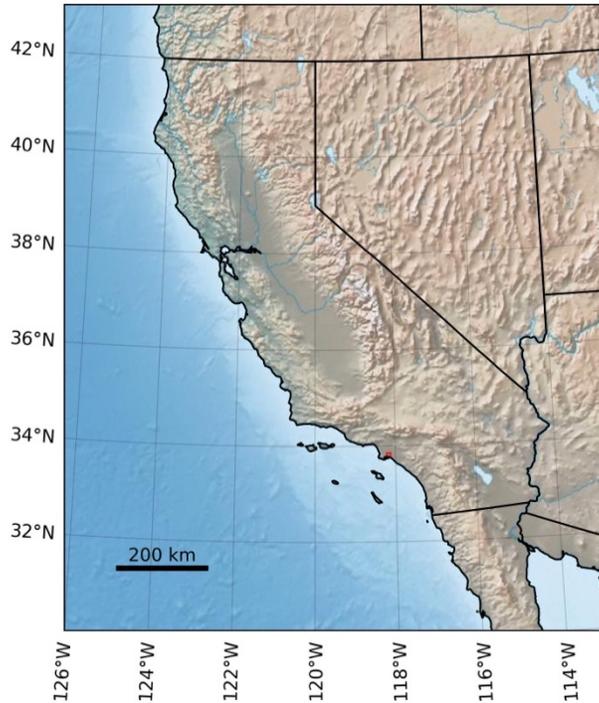

**Figure A25**: The shaded, semi-transparent polygon shows the geographic coverage of the Castellanos LB (Long Beach, California) P-wave speed model listed in Table 4. Background image from Natural Earth (naturalearth.com).

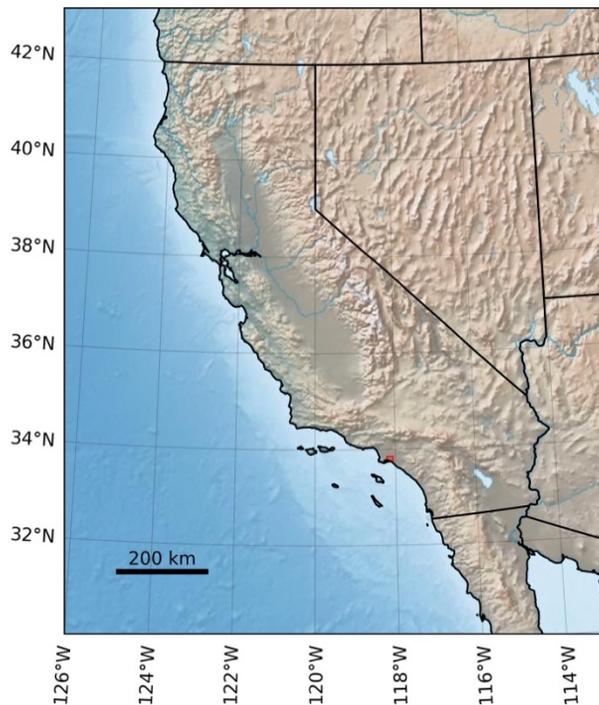

**Figure A26**: The shaded, semi-transparent polygon shows the geographic coverage of the Castellanos LB (Long Beach, California) S-wave speed model listed in Table 4. Background image from Natural Earth (naturalearth.com).

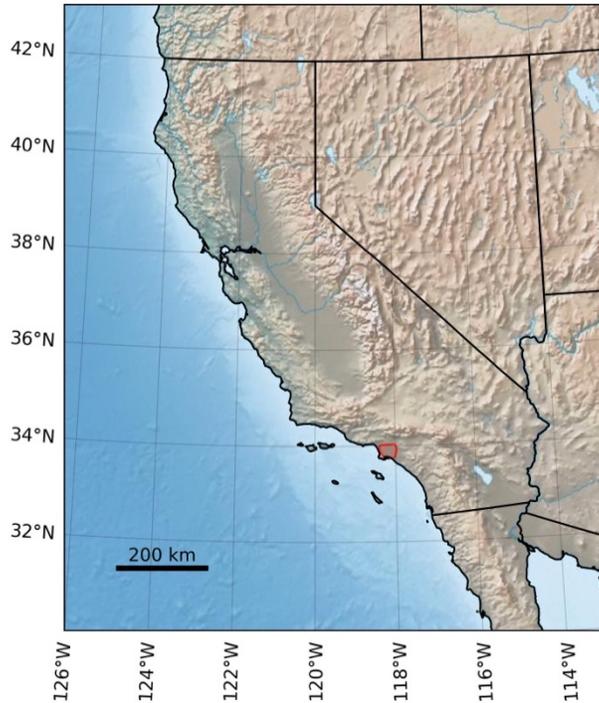

**Figure A27**: The shaded, semi-transparent polygon shows the geographic coverage of the Jia LAS1 (Los Angeles basin) seismic wave speed model listed in Table 4. Background image from Natural Earth (naturalearth.com).

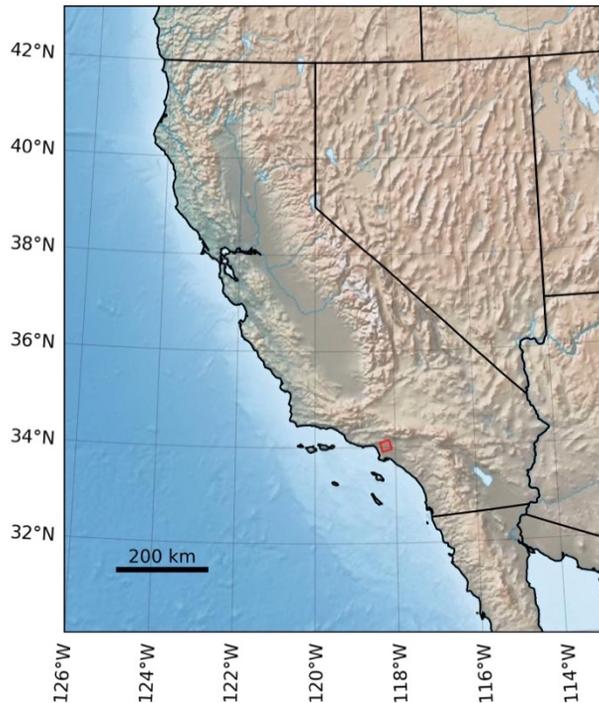

**Figure A28**: The shaded, semi-transparent polygon shows the geographic coverage of the Muir NE LA (northeastern Los Angeles) basin seismic wave speed model listed in Table 4. Background image from Natural Earth (naturalearth.com).

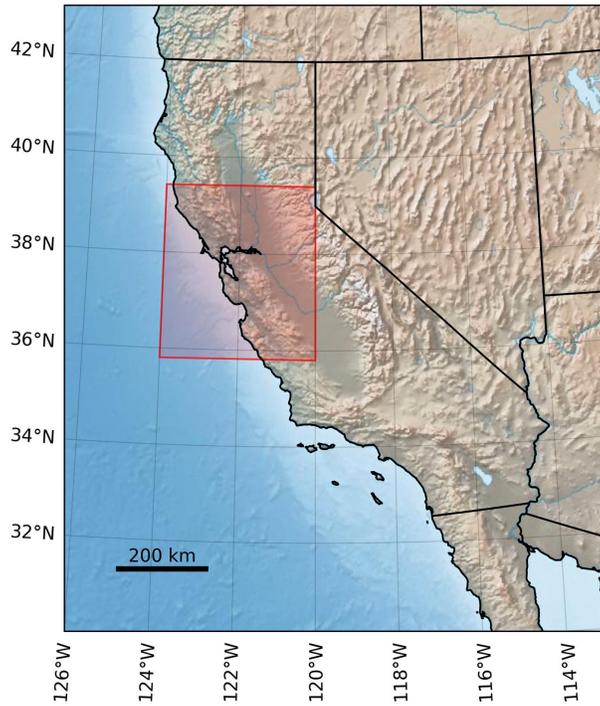

**Figure A29**: The shaded, semi-transparent polygon shows the geographic coverage of the Guo and Thurber seismic wave speed model listed in Table 4. Background image from Natural Earth (naturalearth.com).

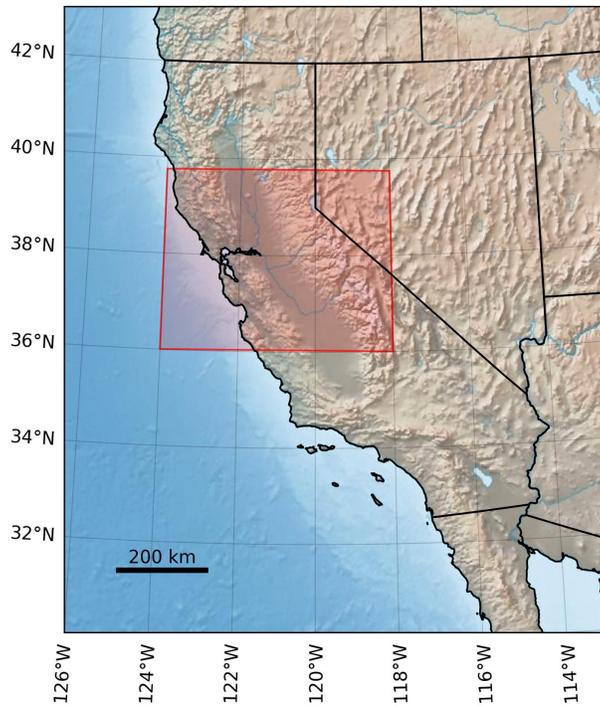

**Figure A30**: The shaded, semi-transparent polygon shows the geographic coverage of the CENOCA_AWT seismic wave speed model listed in Table 4. Background image from Natural Earth (naturalearth.com).

# Rheology models

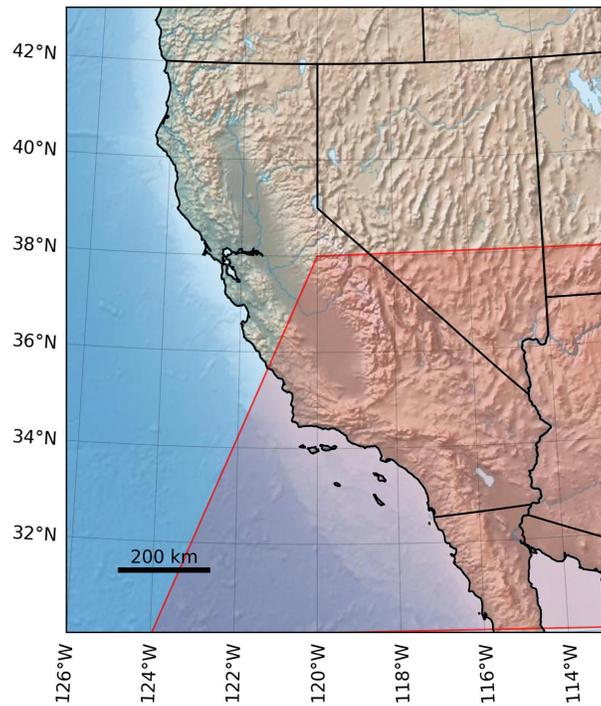

**Figure A31**: The shaded, semi-transparent polygon shows the geographic coverage of the Statewide California Earthquake Center (SCEC) Community Rheology Model (CRM) listed in Table 5. Background image from Natural Earth (naturalearth.com).

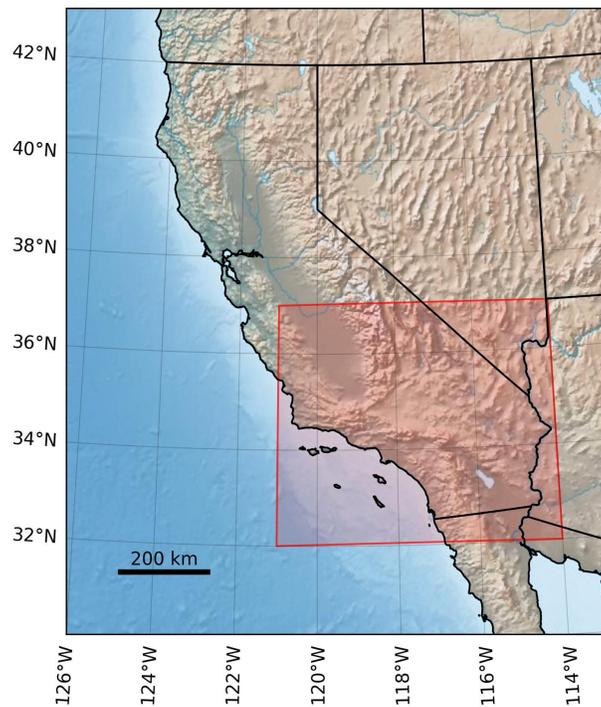

**Figure A32**: The shaded, semi-transparent polygon shows the geographic coverage of the California Thermal Model and Rheology Model (CRM) listed in Table 5. Background image from Natural Earth (naturalearth.com).

# Thermal models

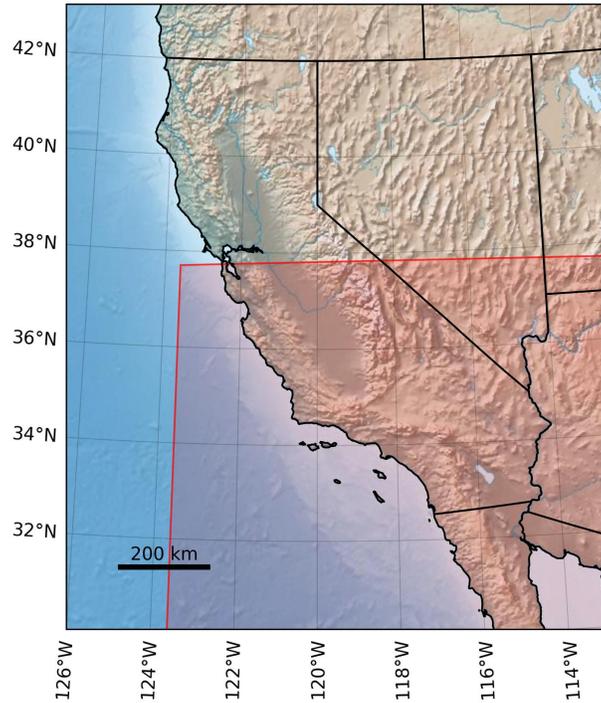

**Figure A33**: The shaded, semi-transparent polygon shows the geographic coverage of the Statewide California Earthquake Center (SCEC) Community Thermal Model (CTM) listed in Table 6. Background image from Natural Earth (naturalearth.com).

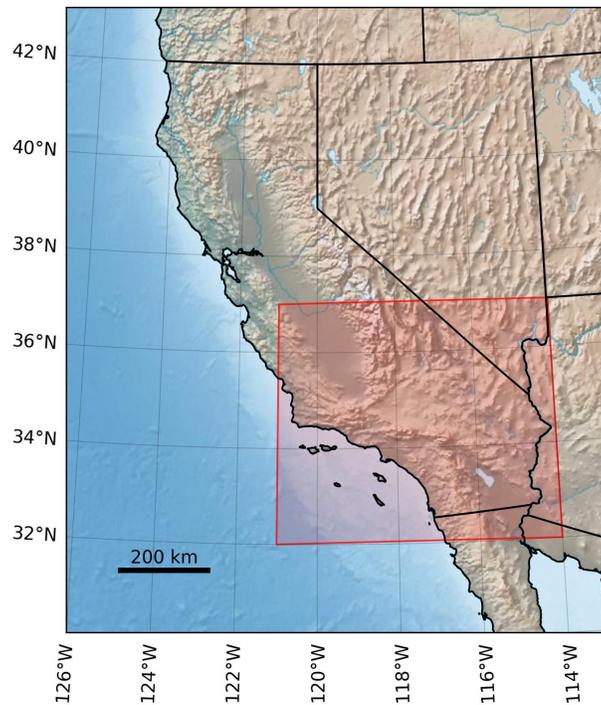

**Figure A34**: The shaded, semi-transparent polygon shows the geographic coverage of the California Thermal Model and Rheology Model listed in Table 6. Background image from Natural Earth (naturalearth.com).

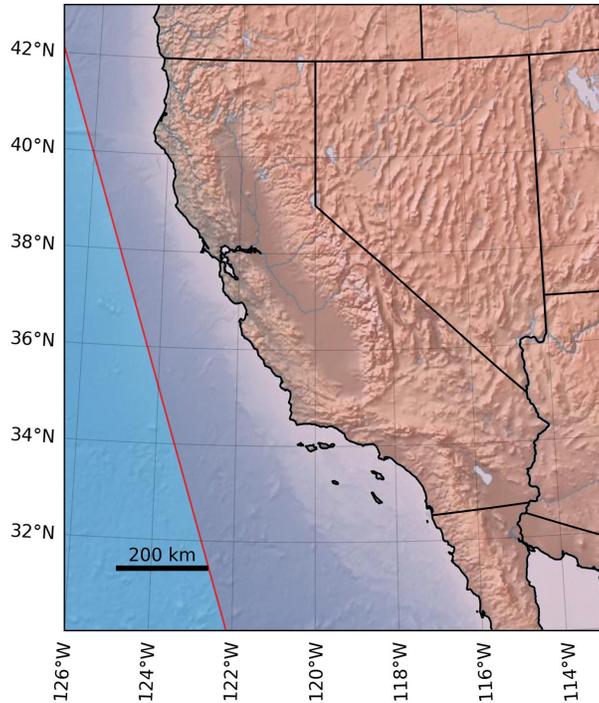

**Figure A35**: The shaded, semi-transparent polygon shows the geographic coverage of the U.S. Geological Survey (USGS) Thermal Model for Seismic Hazard Studies listed in Table 6. Background image from Natural Earth (naturalearth.com).

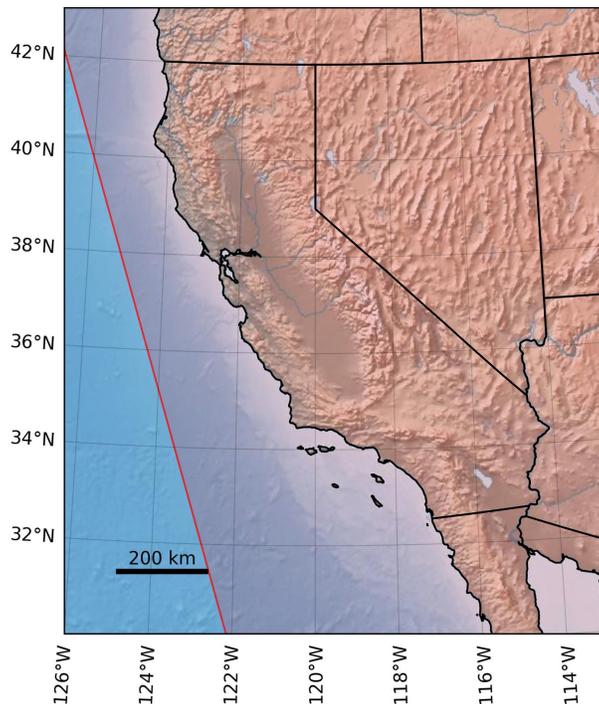

**Figure A36**: The shaded, semi-transparent polygon shows the geographic coverage of the Temperature-At-Depth Maps listed in Table 6. Background image from Natural Earth (naturalearth.com).

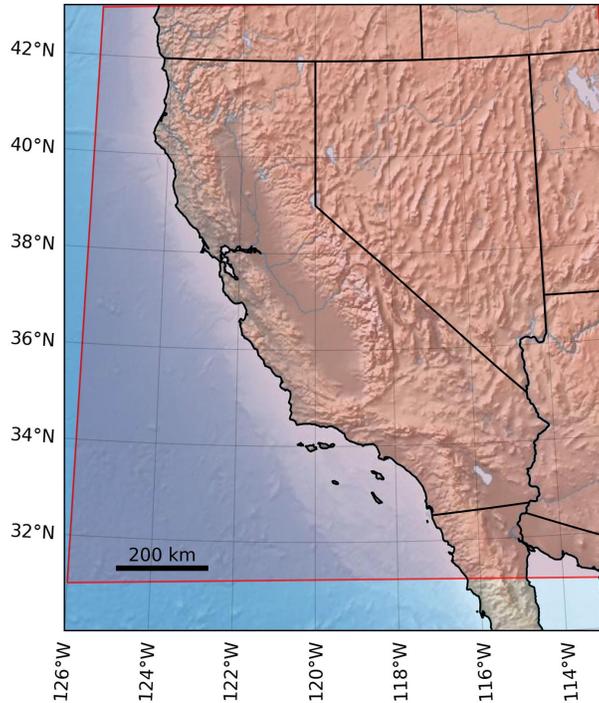

**Figure A37**: The shaded, semi-transparent polygon shows the geographic coverage of the thermal model from Lithospheric Thickness from Sp Receiver Functions listed in Table 6. Background image from Natural Earth (naturalearth.com).

## Stress models

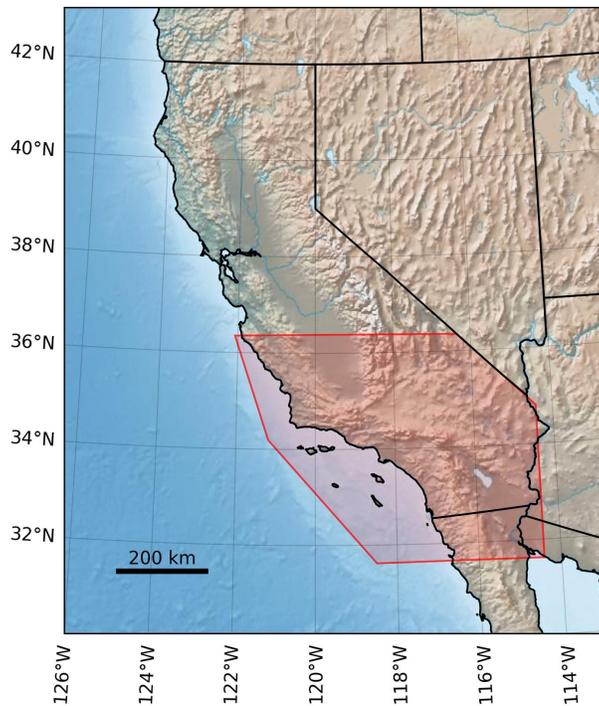

**Figure A38**: The shaded, semi-transparent polygon shows the geographic coverage of the Statewide California Earthquake Center (SCEC) Community Stress Model listed in Table 7. Background image from Natural Earth (naturalearth.com).

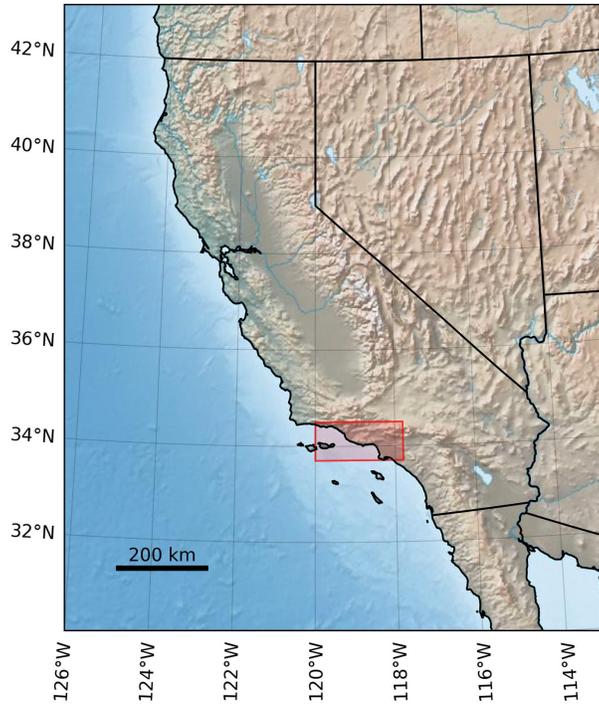

**Figure A39**: The shaded, semi-transparent polygon shows the geographic coverage of the stress model from the Los Angeles Borehole Breakout SHmax listed in Table 7. Background image from Natural Earth (naturalearth.com).

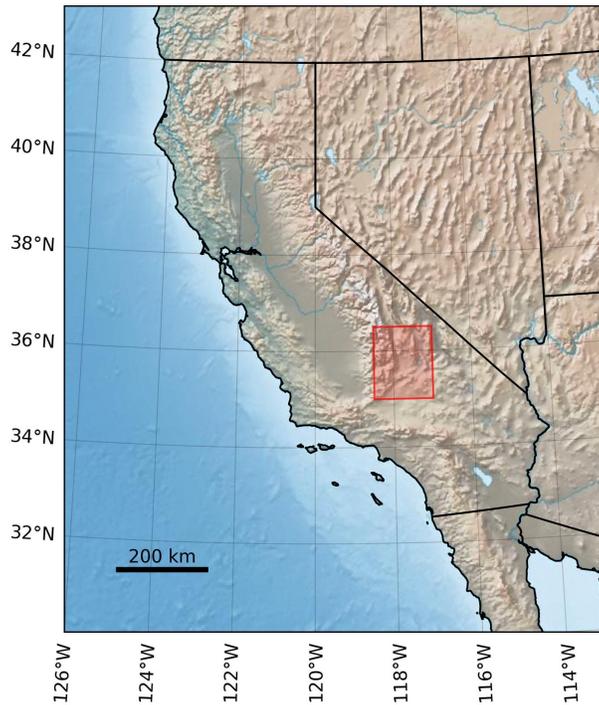

**Figure A40**: The shaded, semi-transparent polygon shows the geographic coverage of the Ridgecrest stress orientation model listed in Table 7. Background image from Natural Earth (naturalearth.com).

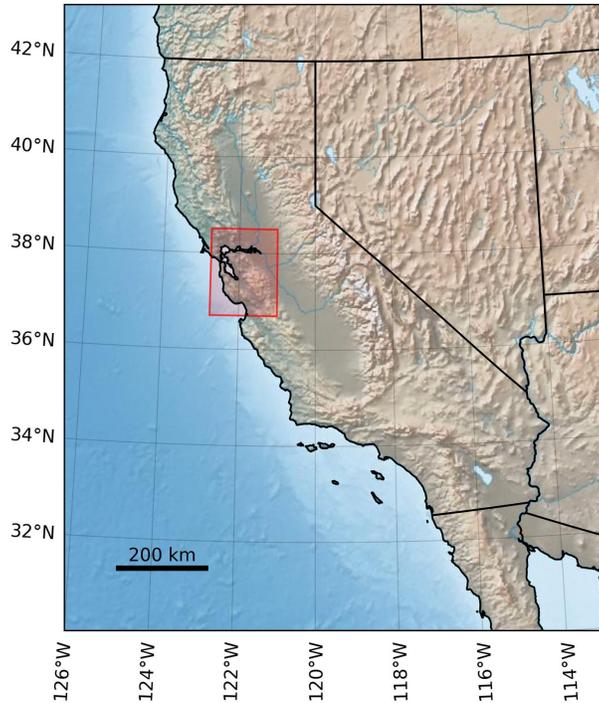

**Figure A41**: The shaded, semi-transparent polygon shows the geographic coverage of the Hardebeck and Michael San Francisco Bay region stress orientation model listed in Table 7. Background image from Natural Earth (naturalearth.com).

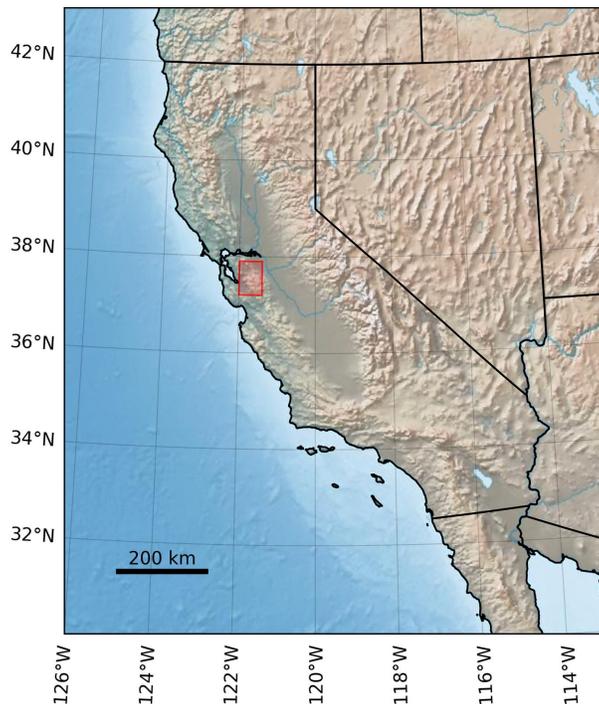

**Figure A42**: The shaded, semi-transparent polygon shows the geographic coverage of the Skoumal and others San Franciso Bay region stress orientation model listed in Table 7. Background image from Natural Earth (naturalearth.com).

# Geodetic models

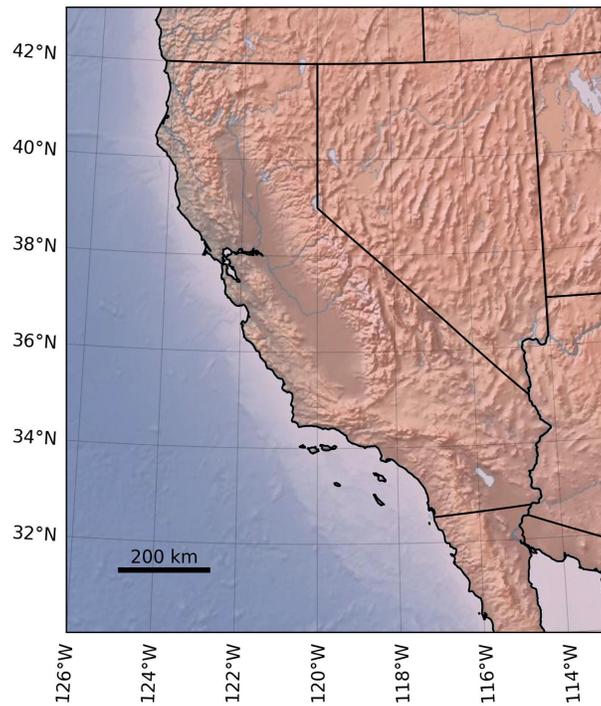

**Figure A43**: The shaded, semi-transparent polygon shows the geographic coverage of the National Seismic Hazard Model (NSHM) velocity field (which covers the entire map) listed in Table 8. Background image from Natural Earth (naturalearth.com).

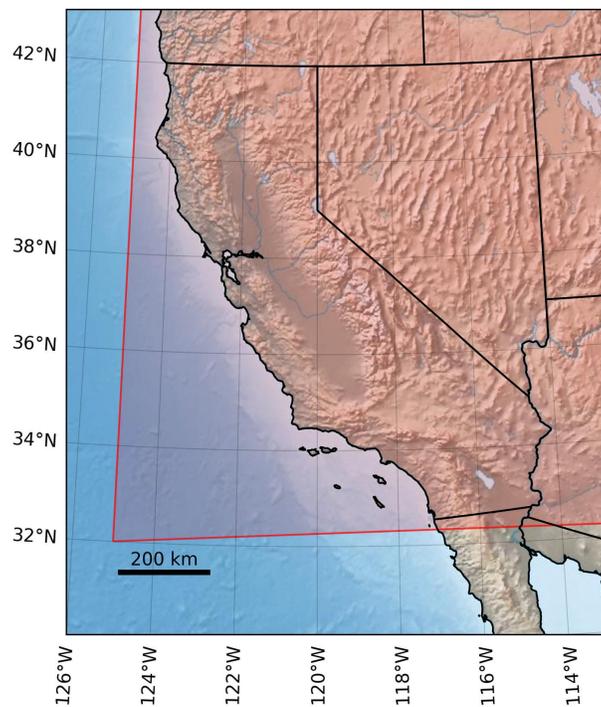

**Figure A44**: The shaded, semi-transparent polygon shows the geographic coverage of the U.S. Geological Survey (USGS) geodetic model from the Global Navigation Satellite System (GNSS) listed in Table 8. Background image from Natural Earth (naturalearth.com).

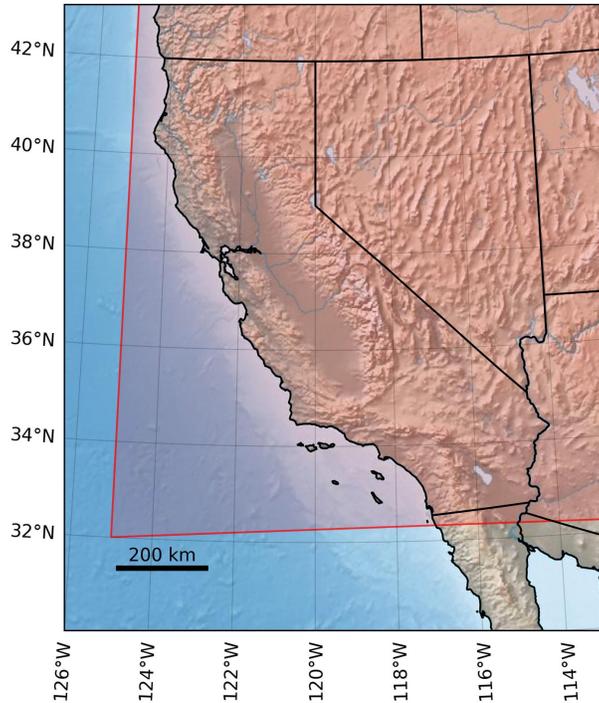

**Figure A45**: The shaded, semi-transparent polygon shows the geographic coverage of the geodetic model from the Advanced Rapid Imaging and Analysis (ARIA) and Observational Products for End-Users from Remote Sensing Analysis (OPERA) listed in Table 8. Background image from Natural Earth (naturalearth.com).

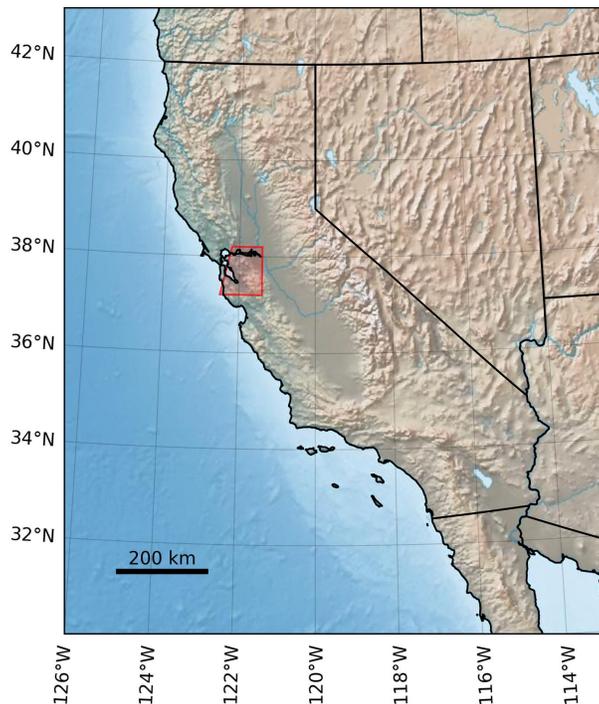

**Figure A46**: The shaded, semi-transparent polygon shows the geographic coverage of the Uninhabited Aerial Vehicle Synthetic Aperture Radar (UAVSAR) geodetic displacement dataset listed in Table 8. Background image from Natural Earth (naturalearth.com).

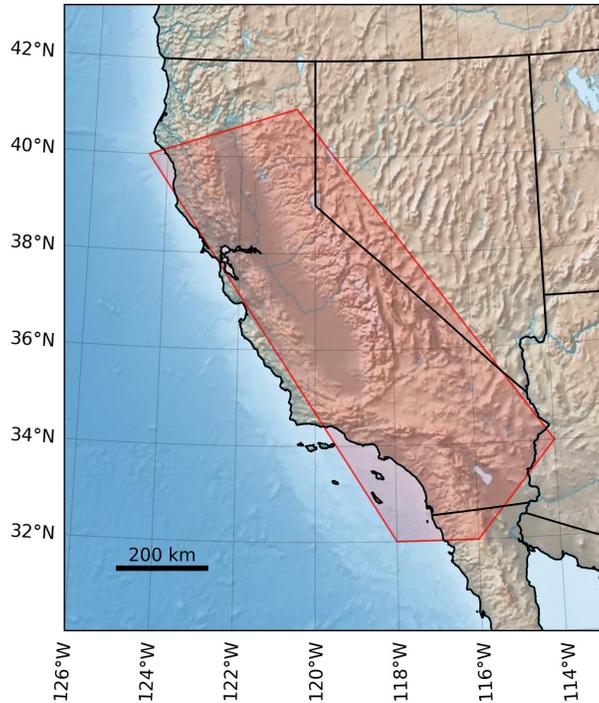

**Figure A47**: The shaded, semi-transparent polygon shows the geographic coverage of the Sentinel01 velocity and time series product listed in Table 8. Background image from Natural Earth (naturalearth.com).

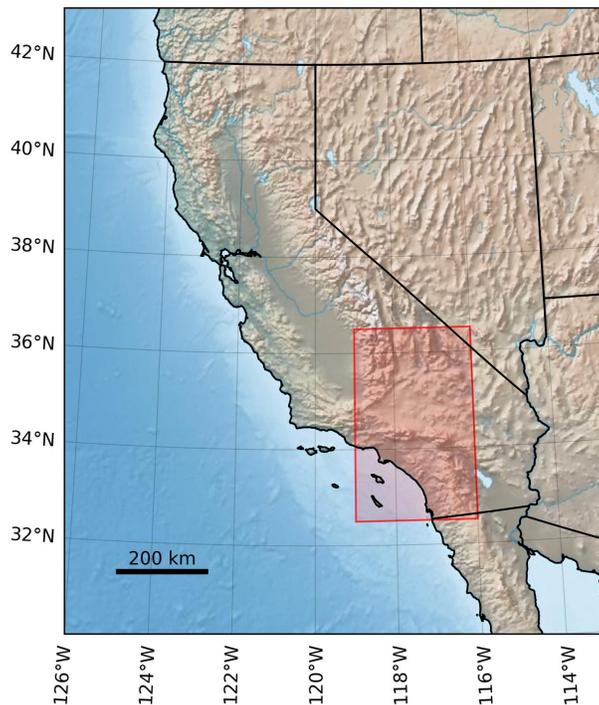

**Figure A48**: The shaded, semi-transparent polygon shows the geographic coverage of the Interferometric Synthetic Aperture Radar (InSAR) and Global Navigation Satellite System (GNSS) combined three-dimensional velocity field listed in Table 8. Background image from Natural Earth (naturalearth.com).

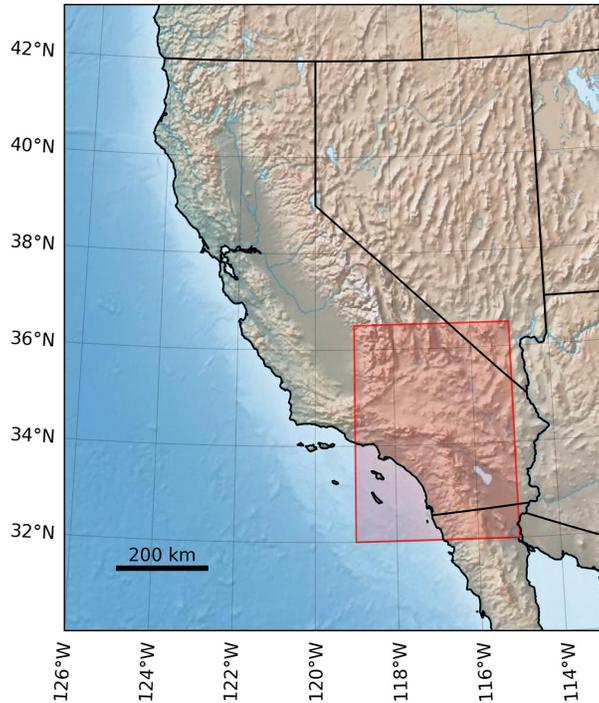

**Figure A49**: The shaded, semi-transparent polygon shows the geographic coverage of the Statewide California Earthquake Center (SCEC) Community Geodetic Model v2.0 listed in Table 8. Background image from Natural Earth (naturalearth.com).

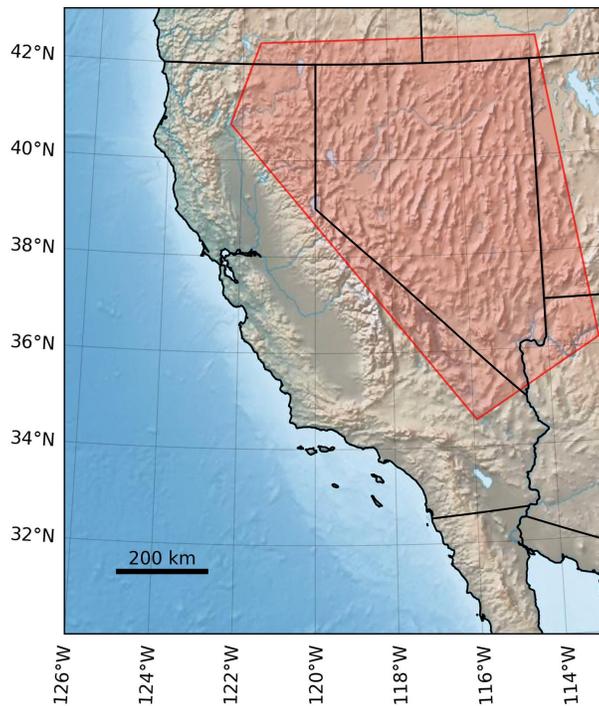

**Figure A50**: The shaded, semi-transparent polygon shows the geographic coverage of the Eastern California Slip Rates model listed in Table 7. Background image from Natural Earth (naturalearth.com).

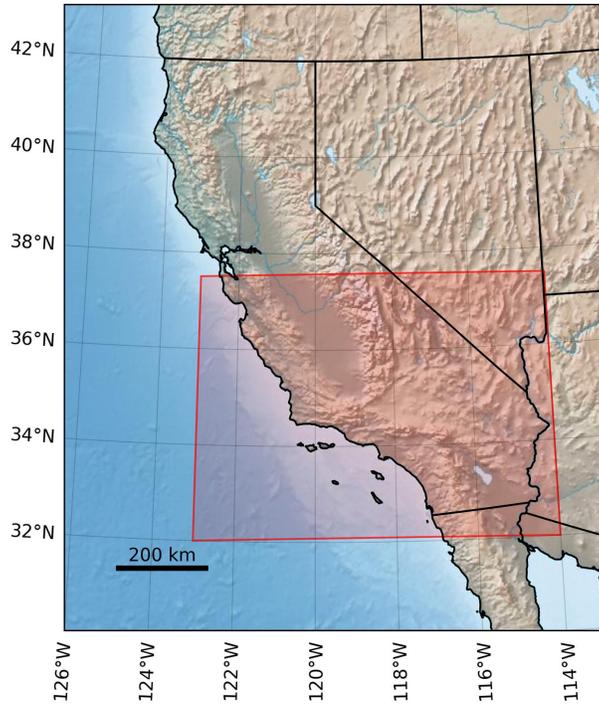

**Figure A51**: The shaded, semi-transparent polygon shows the geographic coverage of the Statewide California Earthquake Center (SCEC) Community Geodetic Model v1.0 strain rate comparison listed in Table 8. Background image from Natural Earth (naturalearth.com).

# Appendix B: Workshop Agenda

Presentation slides may be downloaded by clicking the links following the title. Files are the author's property. They may contain unpublished or preliminary information and should only be used while viewing the talk. Only the presentations for which SCEC has received permission to post publicly are included below.

<u>Monday, March 4, 2024</u>
All times Pacific Time

| | | |
|---|---|---|
| 08:00 - 08:15 | Introduction | Brad Aagaard |
| | Use cases for California community models | |
| 08:15 - 08:35 | What scientific questions could we address with statewide community models? | Alice Gabriel |
| 08:35 - 08:45 | Discussion | |
| 08:45 - 09:05 | How could statewide community models improve seismic hazard assessments? | Christine Goulet |
| 09:05 - 09:15 | Discussion | |
| 09:15 - 09:35 | How could carbon sequestration in California use community models? (PDF, 3.2MB) | Dan Boyd |
| 09:35 - 09:45 | Discussion | |
| 09:45 - 10:00 | Break | |
| | Overview and inventory of existing community models | |
| 10:00 - 10:10 | Geologic models (PDF, 4.9MB) | Russ Graymer / Mike Oskin |
| 10:10 - 10:20 | Fault models (PDF, 6.4MB) | Scott Marshall / Alex Hatem |
| 10:20 - 10:30 | Rheology and thermal models (PDF, 9.0MB) | Laurent Montesi / Wayne Thatcher |
| 10:30 - 10:40 | Stress models (PDF, 2.7MB) | Jeanne Hardebeck / Karen Luttrell |
| 10:40 - 10:50 | Seismic velocity models | Evan Hirakawa / Brad Aagaard |
| 10:50 - 11:00 | Geodetic models (PDF, 2.7MB) | Gareth Funning / Kathryn Materna |
| 11:00 - 11:30 | Discussion | |

<u>Tuesday, March 5, 2024</u>
All times Pacific Time

| | | |
|---|---|---|
| 08:00 - 08:20 | Community models in the Cascadia Region Earthquake Science Center (CRESCENT) (PDF, 3.2MB) | Amanda Thomas |
| 08:20 - 08:30 | Discussion | |
| 08:30 - 08:45 | Breakouts: What does "community" in "community models" mean?<br>1. What are the critical traits for a "community model" to be useful?<br>2. What are the various roles people have in creating, maintaining, and using a "community model"?<br>3. What are important factors for improving "community models"?<br>4. How do we identify, communicate with, and involve community members? | All participants were assigned and divided into the 13 Breakout Groups. |
| 08:45 - 09:15 | Breakout Group Reports (1 slide each, 2-minutes per group) | Breakout Group Leaders |
| 09:15 - 09:30 | Break | |

Techniques for integrating and embedding models

| | | |
|---|---|---|
| 09:30 - 09:40 | Data integration: Geodetic models from GNSS + InSAR (PDF, 1.9MB) | Mike Floyd / Katia Tymofyeyeva |
| 09:40 - 09:50 | Seamless embedding of seismic velocity models via integrated geologic models | Brad Aagaard |
| 09:50 - 10:00 | Embedding high-resolution models in regional models using blending (PDF, 2.5MB) | Patricia Persaud |
| 10:00 - 10:10 | Embedding high-resolution models in regional models using machine learning (PDF, 6.0MB) | Yehuda Ben-Zion |
| 10:10 - 10:20 | Discussion | |
| 10:20 - 10:40 | Plenary discussion: Incentives for participating in community models | |
| 10:40 - 11:00 | Wrap-up discussion: Looking ahead | |

It is SCEC's policy to foster harassment-free environments wherever our science is conducted. By accepting an invitation to participate in a SCEC-supported event, by email or online registration, participants agree to abide by the SCEC Activities Code of Conduct.

# Appendix C: Workshop Participants

A total of 203 participants out of 326 registrants attended the workshop held online using the Zoom platform, with 182 attending on Monday and 150 on Tuesday. The attendees came from both within and outside the United States, with 50% from California, 84% from the United States overall, and 16% internationally, including 7 attendees from Madagascar. Almost 70% of attendees expressed their interest in learning more about community fault and seismic velocity models at the workshop.

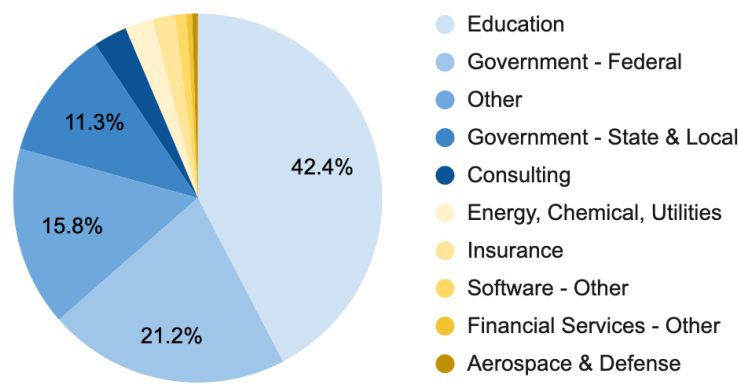

Attendees of the California Community Models Workshop by Industry

Brad Aagaard (USGS)
Rachel Abercrombie (Boston)
Niloufar Abolfathian (OAI)
Esam Abraham (SCEC)
Rasheed Ajala (Columbia)
Young Ho Aladro Chio (CICESE)
Linda Alatik (Linda Alatik Cons)
Richard Allen (UC Berkeley)
Travis Alongi (USGS)
Colin Amos (W Washington)
Titi Anggono (NRIA)
Asif Ashraf (Univ Oregon)
Luciana Astiz (NSF)
Alexis Ault (Utah State)
Sung Bae (Univ Canterbury)
Manochehr Bahavar (EarthScope)
Annemarie Baltay (USGS)
Michael Barall (USGS)
URBI BASU (New Mexico Tech)
Yehuda Ben-Zion (USC)
Scott Bennett (USGS)
Brianna Birkel (USC)
Michael Blanpied (USGS)
Grant Block (Univ New Mexico)
Jacqueline Bott (CGS)
Oliver Boyd (USGS)
Dan Boyd (CGS)
Roland Burgmann (UC Berkeley)
Ashly Cabas (NC State)
Ray Cakir (Washington GS)
Jorge Castellanos (Moody's RMS)
Joanne Chan (USGS)
Rui Chen (CGS)
Xiaowei Chen (Texas A&M)
Robert Clayton (Caltech)
Larry Collins (Cal OES)
James Conrad (USGS)
Tim Dawson (CGS)
Michael DeFrisco (CGS)
Marine Denolle (University of Washington)
Eric Dittmer (Dittmer Consulting)
Mark Dober (AECOM)
Claire Doody (Lawrence Livermore National Laboratory)
Austin Elliott (USGS)
Mariana Eneva (Imageair)
Annde Ewertsen (SSC)
Eric Fielding (JPL, Caltech)
Michael Floyd (MIT)
Gareth Funning (UC Riverside)
Alice Gabriel (UCSD)
Cassie Gann-Phillips (NC State)
Humberto Alfonso García Montano (UNAN-Managua)
Eldon Gath (Earth Consultants)
Eric Geist (USGS)
Farid Ghahari (CGS)
Hadi Ghofrani (Western Univ)
Jianhua Gong (Indiana Univ)
Brad Gooch (CGS)
Christine Goulet (USGS)
alex grant (USGS)
Russell Graymer (USGS)
Katherine Guns (USGS)
Hao Guo (UW Madison)
Bill Hammond (UNR)
Jeanne Hardebeck (USGS)
Behzad Hassani (BC Hydro)
Alex Hatem (USGS)
Egill Hauksson (Caltech)
Elizabeth Hearn (self)
Suzanne Hecker (USGS)
Evan Hirakawa (USGS)
Emilie Hooft (Univ of Oregon)
Yangfan Huang (Oxford)
Tran Huynh (USC/SCEC)
Lorraine Hwang (UC Davis CIG)
Frank Jordan (SBC)
Chun-Yu Ke (Penn State)
Han Kim (Parsons)
Jaehwi Kim (Changwon National)
Hye Jeong Kim (Univ of Utah)
Sangwoo Kim (Moody's RMS)
Chi-Yu King (USGS)
Keith Knudsen (USGS)
Monica Kohler (Caltech)
Folarin Kolawole (Columbia)
Albert Kottke (PG&E)
Fabian Kutschera (UCSD)
Christos Kyriakopoulos (Memphis)
Tyler Ladinsky (CGS)
Martin Lawrence (BC Hydro)
Corinne Layland-Bachmann (LBNL)
Timothy Lee (Indiana)
Isis Lemus (UC Berkeley)

Yuexin Li (Caltech)
Guoliang Li (USC)
Ting Lin (Texas Tech)
Tim Lin (LLNL)
Fan-Chi Lin (Univ of Utah)
Irene Liou (UC Davis)
Zhen Liu (JPL/Caltech)
John Louie (UNR and Terēan)
Julian Lozos (CSUN)
Karen Luttrell (LSU)
Danielle Madugo (CGS)
Chris Madugo (PG&E)
Philip Maechling (SCEC)
Sydney Maguire (Columbia)
Evan Marschall (UC Riverside)
Scott Marshall (App State)
Kathryn Materna (UC Boulder)
Eric Matzel (LLNL)
David Mccallen (LLNL)
Donald Medwedeff (Independent)
Xiaofeng Meng (SCEC)
Christopher Menges (USGS)
Zifei Mi (Donghua University)
Chris Milliner (Caltech)
Sarah Minson (USGS)
Mark Molinari (GeoEngineers)
Laurent Montesi (Univ Maryland)
Angelyn Moore (JPL, Caltech)
Morgan Moschetti (USGS)
Jessica Murray (USGS)
Ayako Nakanishi (JAMSTec)
Yiyu Ni (Univ of Washington)
Craig Nicholson (UCSB)
Tina Niemi (UM-Kansas City)
Zihua Niu (LMU Munich)
Javier Ojeda (Chile & IPGP)
Marc Ollé López (Univ Barcelona)
Evans Onyango (Univ Alaska)
Bar Oryan (UCSD)

Michael Oskin (UC Davis)
Edric Pauk (USC/SCEC)
Patricia Persaud (Univ of Arizona)
Arben Pitarka (LLNL)
Fred Pollitz (USGS)
Manoa Fetra Niaina Rajaonarivelo
Andriniaina Tahina Rakotoarisoa (IOGA)
Tsiriandrimanana Rakotondraibe (IOGA)
Andry Mampionona Ramarolahy (IOGA)
Aaron Rampersad (Beston Consulting)
Manitriniaina Ravoson (IOGA)
Sandra Razafimamonjy (IOGA)
Hoby Razafindrakoto (IOGA)
Gustavo Redondo (Servicio Geol Colombiano)
Tabor Reedy (USBR)
John Rekoske (UCSD)
Arthur Rodgers (LLNL)
Yufang Rong (FM Global)
Anne Rosinski (FEMA)
Stephanie Ross (USGS)
Badie Rowshandel (CEA)
Mousumi Roy (Univ New Mexico)
John Rundle (UC Davis)
Valerie Sahakian (Univ Oregon)
Alexandra Sarmiento (UCLA)
Jeanne Sauber (NASA GSFC)
Frederic Schienberg (UCLA)
David Schwartz (USGS)
Hannu Seebeck (GNS Science)
Israporn Sethanant (Univ Victoria)
Emel Seyhan (Moodys RMS)
Dinesh Shah (NYCDOT Bridges)
Beth Shallon (UC Riverside)
Zheng-Kang Shen (UCLA)

Shuzhong Sheng (ECUT)
William Shinevar (UC Boulder)
Mark Simons (Caltech)
Drake Singleton (USGS)
Angela Stallone (INGV)
Kathleen Steinbroner (CEA)
William Stephenson (USGS)
Joann Stock (Caltech)
IAN STONE (USGS)
Mei-Hui Su (SCEC)
Nigar Sultana (UBC)
Yu-Sheng Sun (U of Oregon)
Karen SUNG (UC Berkeley)
Brian Swanson (CGS)
Taka'aki Taira (UC Berkeley)
Houjun Tang (Berkeley Lab)
Fabia Terra (Berkeley Seismo)
Ensie Teymouri (Memphis)
Wayne Thatcher (USGS)
Amanda Thomas (Univ Oregon)
Cliff Thurber (UW Madison)
Kirk Townsend (CGS)
David Trench (DWR)
Jay Tung (Texas Tech)
Michael Turner (CEC
Russ Van Dissen (GNS Science)
Jessica Velasquez (Moody's)
Yongfei Wang (Verisk)
Ethan Williams (U of Washington)
Erin Wirth (USGS)
Renyi Xu (Donghua University)
Alan Yong (USGS)
Judy Zachariasen (CGS)
Molly Zebker (UT Austin)
Olaf Zielke (KAUST)
倩茹 王 (Donghua University)
璐 陈璐 (Donghua University)